\documentclass[10pt,a4paper,final]{iopart}
\expandafter\let\csname equation*\endcsname\relax
\expandafter\let\csname endequation*\endcsname\relax
\usepackage[margin=1 in]{geometry}
\usepackage{amsmath,color,iopams,graphicx}
\usepackage[breaklinks=true,colorlinks=true,linkcolor=blue,urlcolor=blue,citecolor=blue]{hyperref}
\usepackage[english]{babel}
\usepackage{graphicx,comment}
\usepackage{mathtools}
\usepackage{amsmath}
\usepackage{footnote}
\makesavenoteenv{tabular}
\makesavenoteenv{table}
\usepackage{multicols}
\usepackage{subfigure}
\usepackage{widetext}
\usepackage{caption}
\usepackage{float}
\begin{document}
\title{Non-Markovian transients in transport across chiral quantum wires using space-time non-equilibrium Green functions}
\author{  Nikhil Danny Babu$^{\dagger}$, Girish S. Setlur$^{*}$}
\address{Department of Physics \\ Indian Institute of Technology  Guwahati \\ Guwahati, Assam 781039, India}
\ead{$^{\dagger}$danny@iitg.ac.in,$^{*}$gsetlur@iitg.ernet.in}

\begin{abstract}
We study a system of two non-interacting quantum wires with fermions of opposite chirality with a point contact junction at the origin across which tunneling can take place when an arbitrary time-dependent bias between the wires is applied. We obtain the exact dynamical non-equilibrium Green function by solving Dyson's equation analytically. Both the space-time dependent two and four-point functions are written down in a closed form in terms of simple functions of position and time. This allows us to obtain, among other things, the I-V characteristics for an arbitrary time-dependent bias. Our method is a superior alternative to competing approaches to non-equilibrium as we are able to account for transient phenomena as well as the steady state. We study the approach to steady state by computing the time evolution of the equal-time one-particle Green function. Our method can be easily applied to the problem of a double barrier contact whose internal properties can be adjusted to induce resonant tunneling leading to a conductance maximum. We then consider the case of a finite bandwidth in the point contact and calculate the non-equilibrium transport properties which exhibit non-Markovian behaviour. When a subsequently constant bias is suddenly switched on, the current shows a transient build up before approaching its steady state value in contrast to the infinite bandwidth case. This transient property is consistent with numerical simulations of lattice systems using time-dependent DMRG (tDMRG) suggesting thereby that this transient build up is merely due to the presence of a short distance cutoff in the problem description and not on the other details.  
\end{abstract}

\vspace{2pc}
\section{Introduction}
\begin{figure}
\centering
 \includegraphics[scale=0.7]{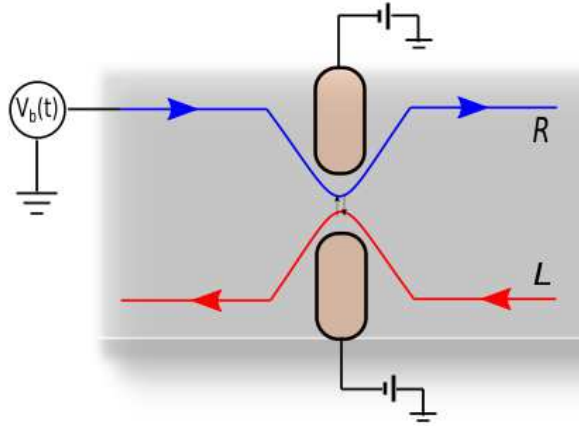}
 \caption{\small Schematic diagram of a tunneling point-contact between two chiral (unidirectional) quantum wires labelled $R$ and $L$. An arbitrary time dependent potential $V_{b}(t)$ is applied on the $R$ branch. A point contact is formed by applying an electrostatic gate voltage.}
 \label{fig1}
 \end{figure}

\noindent It has been over half a century since pioneering theoretical insights shed light on the physics of electron-electron interactions in one-dimensional quantum systems. It was found that a Fermi-liquid description of a 1D electron gas would be destabilised even for a weak repulsive interaction. Later Haldane \cite{haldane1981effective} coined the term ``Luttinger Liquid" to describe the generic state of a one dimensional system of interacting electrons. With advancement in experimental techniques, it has become possible to realise one-dimensional electron systems in semiconductor hetero-structures \cite{meirav1989one,altshuler2012mesoscopic} and recently in carbon nanotubes \cite{bockrath1999luttinger,kim2007tomonaga}. However, the presence of impurities in such experimentally fabricated samples make it very difficult to look for Luttinger Liquid (LL) properties in the lab. This sparked a huge theoretical interest in the problem of impurities in LLs. Kane and Fisher in their pioneering works \cite{kane1992transport,kane1992resonant} showed that for repulsive interactions the electrons are always reflected by a weak link and are transmitted even through a strong barrier in the case of attractive interactions. In addition to this, several analytical\cite{grishin2004functional,matveev1993tunneling,samokhin1998lifetime}
and numerical approaches \cite{qin1996impurity,hamamoto2008numerical,freyn2011numerical,ejima2009luttinger,moon1993resonant}
 have made their mark in this topic.  \\
 One-dimensional chiral Luttinger liquids are realised as the edge states of fractional quantum Hall fluids.
 By using contacts placed at the opposite sides of a tunnel barrier the transport properties related to electron tunneling into fractional quantum hall edges (FQH) with a potential bias between them can be measured. The application of a potential bias drives the system out of equilibrium. The study of non-equilibrium dynamics of the particles is crucial to understand the transport properties. Wen \cite{wen1991edge} predicted a universal scaling form for the tunneling behaviour as a function of bias voltage $V$ and temperature $T$ using a perturbative approach. When the energy scale is determined by the thermal energy the prediction is that the tunneling current obeys a power law in temperature $I_{tun} \propto T^{\alpha-1}$ while depending linearly on $V$ and when the energy scale is dominated by $eV$ the current is nonlinear in bias voltage $I_{tun} \propto V^{\alpha}$. The exponent $\alpha$ is determined by the topological properties of the bulk FQH liquid. Alternatively, Kane and Fisher \cite{kane1992resonant,kane1992transmission} used a renormalisation group (RG) analysis to obtain an expression for the tunneling current in the weak-tunneling limit, which exhibits the same limiting behaviours as in Wen's work. They were able to obtain a duality relation between the strong and weak-tunneling limits. Power law behaviour of the I-V ($I \propto V^{\alpha}$) characteristics was observed for a continuum of filling fractions $\nu$ \cite{chang2003chiral,chang1996observation}. The $\nu=1$ quantum Hall edge states are realized as non-interacting chiral quantum wires. Experimental studies involving one-dimensional constrictions defined by split-gates similar to the schematic shown in Fig.\ref{fig1} have been of interest since early days of quantum transport research \cite{PhysRevB.48.8840,PhysRevB.56.7477,PhysRevB.58.4846,PhysRevLett.61.2797,PhysRevLett.61.2801,PhysRevLett.62.2523,PhysRevLett.76.2145,PhysRevLett.77.135}

 Even when inter-particle interactions are absent, investigation of non-equilibrium dynamics is a non-trivial task. Experimental techniques have advanced in recent years to observe ultrafast transient responses in nanoscale electronic devices. Also experimental methods using tunneling spectroscopy have assisted researchers in understanding the non-equilibrium dynamics of low-dimensional systems better \cite{PhysRevLett.102.036804,altimiras2010non}. Computational Wigner-function approach has been used in the past to study steady state as well as transient regime in a resonant tunneling diode \cite{PhysRevB.39.7720}. But the theoretical tool of choice to study the subtle physics of non-equilibrium systems is the non-equilibrium Green function (NEGF) method (also known as Keldysh formalism). Although there are several techniques such as reduced density matrix methods \cite{bonitz2016quantum}, time-dependent DFT \cite{casida2009time,burke2005time}, quantum master equation \cite{harbola2006quantum}, density matrix renormalisation group \cite{cazalilla2002time,daley2004time,white2004real}, quantum Langevin equation \cite{PhysRevB.86.094503,PhysRevB.75.195110}, Bohm trajectories \cite{leavens1993bohm,oriols2007quantum}, Wigner transport equation \cite{frensley1987wigner,buot1990lattice,agarwal1991exact}, none are well suited to naturally include interactions in the system. The Keldysh formalism to handle the non-equilibrium Green function (NEGF) is a popular choice \cite{keldysh1965ionization,rammer1986quantum}. This method has been extensively used to study tunneling conductance in generic tight binding junctions \cite{berthod2011tunneling}, nonequilibrium currents in quantum dots \cite{doyon2006universal} as well as time dependent transport in double barrier resonant tunneling systems \cite{jauho1994time}.
 \\
 In this work, we explore the problem of non-equilibrium transport across a point-like contact between two chiral (i.e. unidirectional) non-interacting one-dimensional quantum wires (see Fig.\ref{fig1}). In \cite{PhysRevLett.105.156802} the authors consider tunneling between interacting chiral quantum wires albeit treating the point contact perturbatively. Since we treat the point contact exactly, our work complements this earlier work \cite{PhysRevLett.105.156802} and contributes to a comprehensive understanding of this model. We exactly solve for the non-equilibrium Green functions (NEGF) in real space and time from the equation of motion of the Fermi fields as opposed to Fourier transformed energy domain. We extend our analysis beyond the infinite bandwidth limit and consider the case of a finite bandwidth in the point-contact. We obtain a transient in the tunneling current and this term involves an integral over the past history of the system and hence exhibits non-Markovian dynamics. Non-Markovian effects in quantum transport in nanodevices has attracted much attention in recent years over potential applications in quantum computing and nanotechnology. The non-Markovian master equation approach \cite{PhysRevB.83.115439}, analyses based on Keldysh non-equilibrium Green function formalism \cite{jauho1994time,PhysRevB.74.085324,PhysRevB.78.235110,PhysRevB.77.075302} and theories based on the Feynman-Vernon influence functional \cite{Jin_2010} have proven to be very useful in studying transient transport beyond the infinite bandwidth limit. In this work we treat the infinite bandwidth case to be the zeroth order result of a systematic perturbation approach and consider the effects of a finite bandwidth in the point-contact upto first order.\\
  This paper is organised as follows. In Section \ref{sec1} we describe the model Hamiltonian and show in detail the method employed to calculate the finite temperature NEGF in terms of simple functions of position and time in presence of an arbitrary time-dependent voltage bias. In Section \ref{sec2} we calculate the tunneling current and differential tunneling conductance for the non-interacting system in presence of an arbitrary time-dependent voltage bias using the NEGF. In Section \ref{sec3} we point out certain unique non-equilibrium features of this particular model we are studying. We choose to call them \textit{bias-induced anomalies}. In Section \ref{sec4} we compute the dynamical density of states (DDOS). In Section \ref{sec6} we discuss the case of a complex time-dependent tunneling parameter and calculate the tunneling current for this case and show under what circumstances this approach is equivalent to the previously discussed case. We show in Section \ref{db} how our method can be easily applied to the case of resonant tunneling through a double barrier. We then introduce in Section \ref{secfinite} a finite bandwidth in the point-contact and develop a systematic perturbative approach to calculate the transient transport properties which are non-Markovian in this case. Finally in Section \ref{conc} we summarise our main results and its implications.\\
  Exact analyses on time-dependent transport properties in out of equilibrium systems have been done by other authors that go beyond the infinite bandwidth limit \cite{PhysRevB.74.085324}. However in our approach, we analytically obtain the exact non-equilibrium Green functions in space and time domain (two space-time points) as opposed to these other approaches that only consider constant bias and obtain the Green functions in a Fourier transformed frequency domain.  When the point-contact has a finite bandwidth, closed formulas for the full NEGF are possible only as a perturbative series in the inverse of the bandwidth. It is remarkable indeed that a closed formula for the out-of-equilibrium space-time Green function with a general bias of a problem as important as the present one has not been written down till now. This formula provides a convincing derivation of the I-V characteristics that is able to account for a variety of situations such as, time-dependent bias and nonlinearities in the I-V characteristics and so on. Given these observations, it is hard to overstate the importance of the present work.
  \\ \mbox{ } \\ \noindent
  \hyperref[AppendixA]{Appendix A}
  includes some details on how to calculate the tunneling current from the Green functions. \hyperref[AppendixB]{Appendix B} computes the local dynamical density of states in presence of an arbitrary time-dependent bias. \hyperref[AppendixC]{Appendix C} derives the connection between the present formalism and the conventional Keldysh Green functions. Finally, \hyperref[AppendixD]{Appendix D} shows details of the calculation of tunneling current when a finite bandwidth in the point-contact is introduced.

\section{Model and formalism}
\label{sec1}
The tunneling through a point-contact is described by the addition of a tunneling Hamiltonian \cite{PhysRevLett.8.316}. The full Hamiltonian for the system under consideration is,
\begin{align}
H = \sum_{p}(v_Fp+e V_{b}(t) )c^{\dagger}_{p,R}c_{p,R} &+ \sum_{p}(-v_Fp)c^{\dagger}_{p,L}c_{p,L}+ \frac{\Gamma}{L}(c^{\dagger}_{.,R}c_{.,L}+c^{\dagger}_{.,L}c_{.,R})
 \label{eqham}
\end{align}
where $R$ and $L$ label the right and left moving chiral spinless modes and $V_{b}(t)$ is the generic time dependent bias voltage applied to one of the contacts , the right($R$) moving one in this case. $c^{\dagger}_{p}$ and $c_{p}$ are the spinless fermion creation and annihilation operators in momentum space and we use the notation $c^{\dagger}_{.,R} = \sum_{p}c^{\dagger}_{p,R}$ and $\Gamma = \Gamma^{*}$ is the tunneling amplitude of the symmetric point-contact junction and $L$ that does not appear in the subscript is the system size.
 {\it{ Note that a regularisation scheme is implicit when writing down Eq.\ref{eqham}. We call this Dirichlet's regularisation.  Here summing over all momenta implicitly means summing over an interval $ p \in \{-\Lambda,\Lambda\}  $ and then setting $ \Lambda \rightarrow \infty $. This leads us to conclude, among other things, that the values of discontinuous functions are always the average of left and right hand limits. Specifically in the Dirichlet regularisation, the step function that appears repeatedly in the rest of the paper is defined as
 $ \theta(x > 0) = 1, \theta(x < 0 ) = 0, \theta(x=0) = \frac{1}{2} $. }}   We postulate that the right mover experiences a potential $ V_b(t) $ more than the left mover at all spatial locations.
 Note that we have considered a generic time-dependent bias and hence our method is not restricted to the case of a dc-bias alone.

\subsection{Equations of Motion}
We may now go ahead and write down the equations of motion for the Fermi fields for the Hamiltonian in Eq.\ref{eqham} and systematically solve them to obtain the position-time non-equilibrium Green's function. The equations of motion are
\begin{align}
i\partial_t  c_{p,R}(t) &=  (v_Fp + e V_b(t)  ) \mbox{   }    c_{p,R}(t)  + \frac{ \Gamma }{L}  \mbox{   }  c_{.,L}(t) \nonumber \\
i\partial_t  c_{p,L}(t)   &= -v_Fp  \mbox{   }    c_{p,L}(t)    + \frac{ \Gamma }{L}    \mbox{   }   c_{.,R}(t) \mbox{ }\mbox{ }\mbox{ }\mbox{ }\mbox{ }\mbox{ }\mbox{ }\mbox{ }\mbox{ }\mbox{ }\mbox{ }\mbox{ }
\label{eqom}
\end{align}
On solving the above coupled differential equations we write down the solutions explicitly in real space and time
 (where $ \psi_{R/L}(x) = \frac{1}{\sqrt{L}} \sum_p e^{ipx } \mbox{    } c_{p,R/L} $).
\begin{widetext}
\begin{align}
\psi_{R}(x,t)\mbox{        } = \mbox{          }
 U(t_0,t) \left[1
 +2 v_F \Gamma    \mbox{          }  sgn(t_{0}-t + \frac{x}{v_F}) \xi(x,t)\right]\mbox{          }  \psi_{R}(x-v_F(t -t_0),t_0) \nonumber \\
  -i(2 v_F)^2
 U(t - \frac{x}{v_F},t)   \mbox{          }\xi(x,t)\mbox{     }  \psi_{L}(-x+v_F(t -t_0),t_0)
\label{eqR}
 \end{align}
 and
 \begin{align}
\psi_{L}(-x,t)\mbox{        } = \mbox{          } \left[ 1
+2 v_F \Gamma \mbox{    }
 sgn(t_0-t  + \frac{x}{v_F})
 \mbox{ }\xi(x,t) \right]\mbox{ }
\psi_{L}(-x+v_F(t-t_0),t_0)
\nonumber \\
 -i (  2  v_F )^2    \mbox{          }
 U(t_0,t - \frac{x}{v_F} ) \mbox{ }\xi(x,t) \mbox{ } \psi_{R}(x-v_F(t-t_0),t_0)
 \label{eqL}
 \end{align}
 Where we have set,
\begin{align}
\xi(x,t) \equiv  \frac{\Gamma }{v_F}   \mbox{}\frac{ [ \theta(x )  -\theta(v_F t_0 + x- v_F t ) ]}{\Gamma ^2-4 \Gamma ^2 \theta(t - \frac{x}{v_F}-t_{0}) \theta(t_{0}-t + \frac{x}{v_F})+4 v_F^2}
\end{align}
\end{widetext}
and we have defined $U(\tau,t) \equiv e^{ - i \int _{ \tau }^t  e V_{b}(s)ds }$. The time $t_{0}$ specifies, at this stage, some (arbitrary) initial time. Here $\theta(x)$ is the Heaviside step function but regularised using the Dirichlet criterion (i.e. $ \theta(0) = \frac{1}{2} $). Also $ sgn(x) \equiv \theta(x)-\theta(-x) $.
It is easy to verify that the above expressions for the fields do satisfy the equations of motion in real space and time which are
\begin{align}
& i \partial_t \psi_R(x,t)  \mbox{ } = \mbox{ }   ( - i v_F \partial_x + e V_b(t))  \psi_R(x,t)   + \Gamma \mbox{ }\delta(x)\mbox{ }  \psi_L(0,t)
\nonumber \\ &
i \partial_t \psi_L(x,t) \mbox{ } = \mbox{ }     i v_F \partial_x   \psi_L(x,t)  + \Gamma\mbox{  } \delta(x) \mbox{   } \psi_R(0,t)
\end{align}

\subsection{The non-equilibrium Green functions}
We assume that the system is in thermal equilibrium with a reservoir at temperature $ T_{temp} = \frac{1}{\beta}  $ in the remote past since the voltage bias is zero at these early times. Now setting the time $t_{0} = - \infty$ means the same as the system being in the equilibrium state, i.e. before the bias voltage is applied. Using this condition in Eqs.\ref{eqR} and \ref{eqL} we may write,
 \begin{align}
& \psi_{R}(x,t)\mbox{        } = \mbox{          }
 U(-\infty,t) \left[1
-      \theta(x )    \mbox{          }   \frac{  2 \Gamma^2
}{\Gamma ^2 +4 v_F^2}\right]   \psi_{R,0}(x-v_F(t -t_0) < 0,t_0)
 \nonumber\\ &-i  \frac{\Gamma }{v_F}
    \theta(x )   \mbox{          } U(t - \frac{x}{v_F},t)   \mbox{          } \frac{(2 v_F)^2
}{\Gamma ^2 +4 v_F^2}\mbox{     } \psi_{L,0}(-x+v_F(t -t_0) > 0,t_0)
\label{psiR}
 \end{align}
 and
 \begin{align}
& \psi_{L}(-x,t)\mbox{        } = \mbox{          } \left[ 1
-      \theta(x )    \mbox{          }   \frac{  2 \Gamma^2
}{\Gamma ^2 +4 v_F^2}
\right] \psi_{L,0}(-x+v_F(t-t_0) > 0,t_0)\nonumber
\\ &-i \frac{\Gamma }{v_F}     \theta( x )     \mbox{          } \frac{ (  2  v_F )^2    \mbox{          }
 U(-\infty,t - \frac{x}{v_F} ) }{\Gamma ^2 +4 v_F^2}
\mbox{          }\psi_{R,0}(x-v_F(t-t_0) < 0,t_0)
\label{psiL}
 \end{align}
 Before calculating the two-point Green's function it is crucial to note how the point-contact tunneling amplitude is related to the reflection ($R$) and transmission ($T$) amplitudes
 ($ |T|^2 + |R|^2 = 1 $) when modelled as an isolated impurity as shown in the reference \cite{babu2020density},
 \begin{align}
 \Gamma = \Gamma^{*} = \frac{4 i v_{F} R T}{2T^{2} + 2 T}
 \end{align}
 This means,
 \begin{align}
 R \mbox{        } = \mbox{          } -i \frac{4\Gamma    v_F }{\Gamma ^2+4 v_F^2} \mbox{ };\mbox{ }T \mbox{        } = \mbox{          }  \frac{4v_F^2-\Gamma ^2  }{\Gamma ^2+4 v_F^2}
 \label{eqrt}
 \end{align}
 Note that since $t_{0}$ is in the equilibrium region ($t_{0} = -\infty$) the quantities $x-v_F(t-t_0) < 0$ and $-x+v_F(t-t_0) > 0$ for any fixed $ x,t $. The correlation functions at time $t_{0} = -\infty $ then are the well-known finite temperature equilibrium Green functions \cite{das2018quantum}. Set $z \equiv x-v_F(t-t_0),\mbox{   }  z^{'} \equiv x^{'}-v_F(t^{'}-t_0) $,
 \begin{align}
& < \psi^{\dagger}_{R,0}(z^{'}<0,t_0) \psi_{R,0}(z<0,t_0)>\mbox{          } = \mbox{          }
 <\psi^{\dagger}_{L,0}(-z^{'} >0,t_0)\psi_{L,0}(-z>0,t_0)>\mbox{          } = \mbox{          }\nonumber  \\&  - \frac{i}{2\pi} \frac{ \frac{ \pi }{\beta v_F } }{\sinh( \frac{ \pi }{\beta v_F }
 (x-x^{'}-v_F(t-t^{'}) ) )  }
    \label{eqrrz}
 \end{align}
 where $\beta$ is the inverse temperature. The $R,L$  and $L,R$ correlations with the position coordinates at the opposite sides of the point-contact are zero.
 \begin{align}
 &< \psi^{\dagger}_{R,0}(z^{'}<0,t_0)\psi_{L,0}(-z>0,t_0)> \nonumber \\ &=\mbox{ }  <\psi^{\dagger}_{L,0}(-z^{'}>0,t_0)\psi_{R,0}(z<0,t_0)> \mbox{ }=\mbox{ } 0
 \label{eqllz}
 \end{align}
 Using Eqs.\ref{eqrrz}-\ref{eqllz} in Eqs.\ref{psiR}-\ref{psiL} and the corresponding conjugates, we get the full non-equilibrium Green functions (NEGF),
 \begin{align}
 <\psi^{\dagger}_{\nu^{'}}(x^{'},t^{'})\psi_{\nu}(x,t)> \mbox{ }=\mbox{ } -\frac{i}{2\pi} \frac{\frac{\pi}{\beta v_{F}}}{\sinh( \frac{ \pi }{\beta v_F } (\nu x-\nu^{'}x^{'}-v_F(t-t^{'}) ) )} \kappa_{\nu,\nu^{'}}
\label{eqnoneq}
 \end{align}
 where $\nu$,$\nu^{'} = \pm 1$ with $R=1$ and $L=-1$ and
 \begin{widetext}
 \small
 \begin{align}
 \kappa_{1,1} \mbox{ }=\mbox{ } \bigg(  U(t^{'},t) \left[1
-      \theta(x^{'})    \mbox{          }   \frac{  2 \Gamma^2
}{\Gamma ^2 +4 v_F^2}\right]&\mbox{          }    \left[1
-      \theta(x )    \mbox{          }   \frac{  2 \Gamma^2
}{\Gamma ^2 +4 v_F^2}\right]
  \nonumber \\ &+ \left( \frac{\Gamma }{v_F} \mbox{          } \frac{(2 v_F)^2
}{\Gamma ^2 +4 v_F^2}\right)^2 \mbox{          }
  \theta(x )     \theta(x^{'} )   \mbox{          } U(t^{'},t^{'} - \frac{x^{'}}{v_F})   \mbox{     }
   U(t - \frac{x}{v_F},t)   \bigg)
   \label{eqkappa1}
 \end{align}
 \begin{align}
 \kappa_{-1,-1} \mbox{ }=\mbox{ } \bigg( \left[ 1
-      \theta(-x^{'})    \mbox{          }   \frac{  2 \Gamma^2
}{\Gamma ^2 +4 v_F^2}
\right] &  \left[ 1
-      \theta(-x )    \mbox{          }   \frac{  2 \Gamma^2
}{\Gamma ^2 +4 v_F^2}
\right]   \nonumber \\ &+   \left( \frac{\Gamma }{v_F} \mbox{          } \frac{(2 v_F)^2
}{\Gamma ^2 +4 v_F^2}\right)^2 \mbox{          }  \theta(-x )    \theta( -x^{'} )    \mbox{          }
 U(t^{'} + \frac{x^{'}}{v_F},t + \frac{x}{v_F} ) \bigg)
 \label{eqkappa2}
 \end{align}
 \begin{align}
 \kappa_{1,-1} \mbox{ }= \mbox{ }\bigg( -U(t - \frac{x}{v_F},t)   \mbox{          }  &\left[ 1
-      \theta(-x^{'})    \mbox{          }   \frac{  2 \Gamma^2
}{\Gamma ^2 +4 v_F^2}
\right]
    \theta(x )   \mbox{          } \nonumber \\ &+
 U(t^{'} + \frac{x^{'}}{v_F},t)
\mbox{          }
\left[1
-      \theta(x )    \mbox{          }   \frac{  2 \Gamma^2
}{\Gamma ^2 +4 v_F^2}\right] \theta( -x^{'} )   \bigg) i  \frac{\Gamma }{v_F}\frac{(2 v_F)^2
}{\Gamma ^2 +4 v_F^2}
\label{eqkappa3}
 \end{align}
 \begin{align}
 \kappa_{-1,1} \mbox{ }=\mbox{ }\bigg( - U(t^{'},t + \frac{x}{v_F} ) &\left[1
-      \theta(x^{'})    \mbox{          }   \frac{  2 \Gamma^2
}{\Gamma ^2 +4 v_F^2}\right]  \theta( -x )
 \nonumber \\ &+     U(t^{'},t^{'} - \frac{x^{'}}{v_F})   \mbox{          }     \left[ 1
-      \theta(-x )    \mbox{          }   \frac{  2 \Gamma^2
}{\Gamma ^2 +4 v_F^2}
\right]  \theta(x^{'} )     \bigg) \mbox{          } i \frac{\Gamma }{v_F}    \frac{ (  2  v_F )^2   }{\Gamma ^2 +4 v_F^2}
\label{eqkappa4}
 \end{align}
 \normalsize
 \end{widetext}
 \begin{figure*}
 \centering
\subfigure[]{\includegraphics[width=45mm,height=45mm]{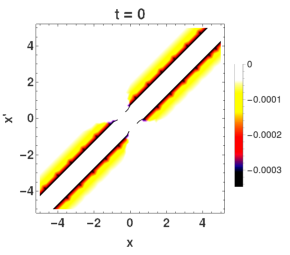}}\quad \quad
\subfigure[]{\includegraphics[width=45mm,height=45mm]{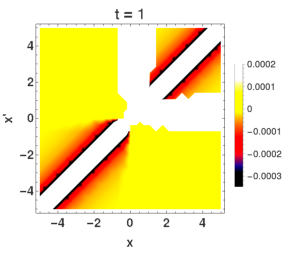}}\quad \quad
\subfigure[]{\includegraphics[width=45mm,height=45mm]{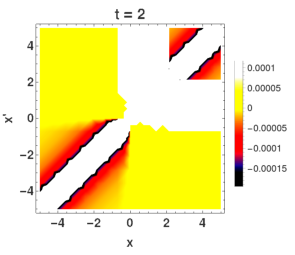}}\\
\subfigure[]{\includegraphics[width=45mm,height=45mm]{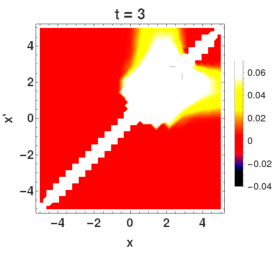}}\quad \quad
\subfigure[]{\includegraphics[width=45mm,height=45mm]{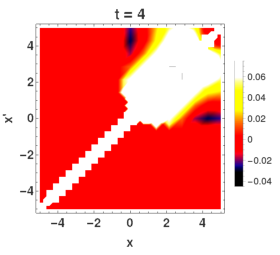}}\quad \quad
\subfigure[]{\includegraphics[width=45mm,height=45mm]{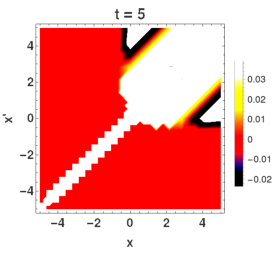}}\\
\subfigure[]{\includegraphics[width=45mm,height=45mm]{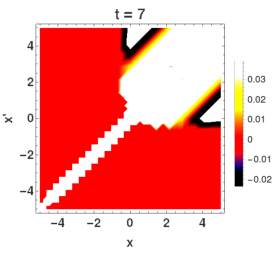}}\quad \quad
\subfigure[]{\includegraphics[width=45mm,height=45mm]{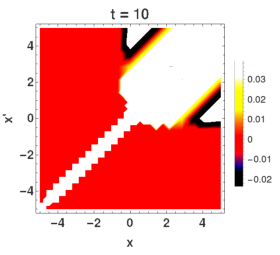}}\quad \quad
\subfigure[]{\includegraphics[width=45mm,height=45mm]{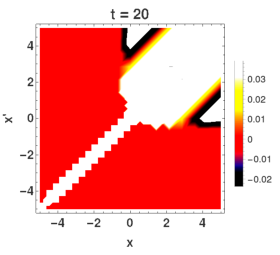}}
\caption{Transient in NEGF: \small This figure shows density plots of the real part of the equal-time RR Green function  $Re[<\psi^{\dagger}_{R}(x^{'},t)\psi_{R}(x,t)>]$  vs time ($t$) in presence of a step bias that is switched on at $t=0$. The real part of the non-equilibrium Green functions shows transient dynamics (\textbf{a})-(\textbf{e}) before reaching a steady state (\textbf{f})-(\textbf{i}). The other parameters are chosen to be $\Gamma=2$ and $v_{F} = 1$.}
\label{gtrans}
\end{figure*}
 For illustrative purposes, let us consider a step bias that is initially zero but is turned on suddenly at $t=0$ and remains constant for all positive times. We can show by plotting the equal-time NEGF versus the spatial coordinates for different times that at small times the NEGF captures the transient behaviour of the system and reaches a steady state at large positive times. In Fig.\ref{gtrans} the zero temperature equal-time RR non-equilibrium Green function for various times is plotted, clearly showing the initial transient regime followed by the appearance of a steady state at sufficiently long times after the switching on of a constant bias. We show in \hyperref[AppendixC]{Appendix C} how the space-time non-equilibrium Green function we have written in Eq.\ref{eqnoneq} can be interpreted as the Keldysh contour ordered NEGF.

 \subsection{Consistency check in the equilibrium limit}
 In a steady-state, the NEGF is a function of the time-difference $t-t^{'}$. The equilibrium Green's function also depends only on the time difference but the non-equilibrium nature of the problem is evident in the fact that the NEGF does not obey the KMS boundary conditions in imaginary time \cite{kubo1957statistical,martin1959theory} even when the system has reached a steady state. It is an important consistency check to make sure that the NEGF reduces to the equilibrium Green's function when the times $t$ and $t^{'}$ are set to be in the equilibrium region (remote past). This is equivalent to setting $U(t^{'},t) =1$ since there is no voltage bias when the system is in equilibrium.
  Note that when $xx^{'} > 0$,
 \begin{align}
    \bigg(   \left[1
-      \theta(x^{'})    \mbox{          }   \frac{  2 \Gamma^2
}{\Gamma ^2 +4 v_F^2}\right]\mbox{          }    \left[1
-      \theta(x )    \mbox{          }   \frac{  2 \Gamma^2
}{\Gamma ^2 +4 v_F^2}\right]
 +   \theta(x )  \theta(x^{'} )   \mbox{          } \left( \frac{\Gamma }{v_F}
    \frac{(2 v_F)^2
}{\Gamma ^2 +4 v_F^2} \right)^2     \mbox{          }
 \bigg) = 1
 \end{align}
 whereas from Eq.\ref{eqrt} it is clear that when $xx^{'} < 0$,
 \begin{align}
    &\bigg(   \left[1
-      \theta(x^{'})    \mbox{          }   \frac{  2 \Gamma^2
}{\Gamma ^2 +4 v_F^2}\right]\mbox{          }    \left[1
-      \theta(x )    \mbox{          }   \frac{  2 \Gamma^2
}{\Gamma ^2 +4 v_F^2}\right]
 +   \theta(x )  \theta(x^{'} )   \mbox{          } \left( \frac{\Gamma }{v_F}
    \frac{(2 v_F)^2
}{\Gamma ^2 +4 v_F^2} \right)^2     \mbox{          }
 \bigg)
\mbox{        } \nonumber \\ &= \mbox{          }        \left[1
-        \frac{  2 \Gamma^2
}{\Gamma ^2 +4 v_F^2}\right]   = T = T^*
 \end{align}
 The NEGF in this case, reduces to the well-known \cite{das2018quantum} equilibrium Green functions given in Eq.\ref{eqg}.  \begin{align}
<T \psi^{\dagger}_{\nu^{'}}(x^{'},t^{'})\psi_{\nu}(x,t) >_{equil}\mbox{        } = \mbox{          }- \sum_{\gamma,\gamma^{'} = \pm 1} \frac{\pi}{\beta v_{F}} \frac{\theta(\gamma x)\theta(\gamma^{'}x^{'} )g_{\gamma,\gamma^{'}}(\nu,\nu^{'})}{\sinh( \frac{ \pi }{\beta v_F } (\nu x-\nu^{'} x^{'}-v_F(t-t^{'}) ) )}
 \label{eqg}
 \end{align}
 where in terms of the reflection and transmission amplitudes
 \begin{align}
 g_{\gamma,\gamma^{'}} (\nu,\nu^{'})\mbox{ } = \mbox{ } \frac{i}{2\pi} \bigg( \delta_{\nu,\nu^{'}} \delta_{\gamma,\gamma^{'}} +(T \delta_{\nu,\nu^{'}}+R \delta_{\nu,-\nu^{'}})\delta_{\gamma,\nu}\delta_{\gamma^{'},-\nu^{'}}+(T^{*} \delta_{\nu,\nu^{'}}+R^{*} \delta_{\nu,-\nu^{'}})\delta_{\gamma,-\nu}\delta_{\gamma^{'},\nu^{'}}\bigg)
 \end{align}
 It is easy to show that the non-equilibrium Green functions satisfy the equation of motion for the two-point functions. Now that we have obtained the NEGF we can use it to study the transport properties of the system particularly the tunneling current and differential tunneling conductance with general time-dependent bias voltage of which the well studied problem of a dc-bias \cite{berthod2011tunneling,ferrer1988contact,chen1991dynamic} is a special case. In the infinite bandwidth case only steady state behaviour in the $I-V$ characteristics is observed. When we consider the case of a finite bandwidth in the point-contact we observe non-Markovian transient transport dynamics in presence of an arbitrary time-dependent bias voltage.
 \section{Tunneling current and conductance in the infinite bandwidth case}
 \label{sec2}
 In this section, we evaluate the tunneling current and differential tunneling conductance. The tunneling current is defined usually as the rate of change of the difference in the number of right and left movers,
 \begin{align}
 I_{tun}(t) = e\mbox{ }\partial_{t}\frac{\Delta N}{2} = e\frac{i}{2}\left[H,\Delta N\right] = e\frac{i}{2}\left[H,N_{R}-N_{L}\right]
 \label{tundef}
 \end{align}
 We use the convention as in \cite{shah2016consistent}, where they have considered $\mu_{L} - \mu_{R} = e V$. In our case the bias is applied to the right movers so that, $\mu_{R} = e V_{b}(t)$ and $\mu_{L} = 0$. Hence in our case we use $\mu_{L}-\mu_{R} = -e V_{b}(t) = e V(t)$. The current becomes,
 \begin{align}
  I_{tun}(t) \mbox{        } = \mbox{          }-i e \Gamma \mbox{        } \lim_{t^{'} \rightarrow t } \bigg( < \psi^{\dagger}_R(0,t^{'})  \psi_L(0,t) > - < \psi^{\dagger}_L(0,t)  \psi_R(0,t^{'}) > \bigg)
  \label{eqitun}
 \end{align}
 The detailed calculation of tunneling current is shown in \hyperref[AppendixA]{Appendix A}. Using the form of the NEGF we have obtained in Eq.\ref{eqitun}, finally we get  ($\hbar = 1$),
 \begin{align}
 I_{tun}(t) \mbox{        } = \mbox{          }   - \Gamma^2 \mbox{        }
     \mbox{          }    \frac{   4   }{\Gamma ^2 +4 v_F^2} \left[1
-       \frac{    \Gamma^2
}{\Gamma ^2 +4 v_F^2}\right]
\mbox{    }
\frac{e^2}{2\pi } V_b(t)
 \end{align}
 We define the tunneling amplitude in terms of a tunneling parameter $t_{p}$ as $\Gamma = 2 v_{F} t_{p}$ \cite{filippone2016tunneling}. So that the tunneling current upon restoring dimensional units and using $V_{b}(t) = -V(t)$
 \begin{align}
 I_{tun}(t) \mbox{        } = \mbox{          }
     \mbox{          }    \frac{   4 t_{p}^{2}  }{(t_{p}^{2} +1)^{2}}
\mbox{    }
\frac{e^2}{h } V(t)
\label{eqtunc}
 \end{align}
 and the differential tunneling conductance
 \begin{align}
 G = \frac{d I_{tun}}{d V(\tau)} \mbox{ }=\mbox{ } G_{tun} \mbox{        } = \mbox{          }  \frac{   4 t_{p}^{2}  }{(t_{p}^{2} +1)^{2}}
\mbox{    }
\frac{e^2}{ h }
\label{eqcond}
 \end{align}
 are obtained as functions of the tunneling parameter and agree with the predictions of standard scattering theory \cite{LANDAUER198191,PhysRevB.31.6207}.
 The tunneling conductance shows $t_{p} \rightarrow \frac{1}{t_{p}}$ duality between the strong and weak tunneling regimes as predicted \cite{filippone2016tunneling,shah2016consistent}. Although our calculations are done when the system was at  a finite temperature in the distant past,  the temperature dependence naturally drops out of the expression for tunneling current when there are no interparticle interactions. Also it is worth mentioning that the current is   linearly dependent on the time-dependent voltage bias (linear response) and the nonlinear dependence arises only when interactions between the fermions are taken into account (which is not done in the present paper).\\
 For a step bias it is clear from Eq.\ref{eqtunc} that for the case of an infinite bandwidth no transients appear in the tunneling current even though the NEGF (Eq.\ref{eqnoneq}) shows transients. But these transients drop out when one evaluates the equal space-equal time two-point functions at the origin such as $< \psi^{\dagger}_R(0,t)  \psi_L(0,t) >$ and $< \psi^{\dagger}_L(0,t)  \psi_R(0,t) >$ that are present in the expression for the tunneling current (Eq.\ref{eqitun}). This could be explained by thinking of the variable $|x - x^{'}|$ as serving as a length scale or inverse bandwidth (for $< \psi^{\dagger}_R(x,t)  \psi_R(x^{'},t) > $ and $< \psi^{\dagger}_L(x,t)  \psi_L(x^{'},t) > $). This proxy for a inverse bandwidth is present in the full NEGF but is zero when one evaluates the tunneling current (because in this case $x = x^{'}$). So in order to investigate transients in the tunneling current one has to introduce a finite bandwidth in the problem description explicitly which is what we do in Sec.\ref{secfinite}. In that case one has to forego the idea of an exact solution and resort to a systematic perturbative approach.  
  \section{Bias-induced anomalies}
 \label{sec3}
 In this section, we point out two interesting features of the problem we are studying. They are caused by the presence of a bias in the system which leads to physical quantities that normally vanish identically to be non-zero.

 Specifically, we contrast the behavior of a) the time-rate of change of the density difference between the local right and left mover densities and b) the time-rate of change of the difference between the total number of right movers and left movers.
\\ \mbox{ } \\ The quantity in a) is,
 \begin{align}
\partial_t \Delta \rho(x,t)  \mbox{        } = \mbox{          }    \frac{d}{dt}  (\rho_R(x,t) - \rho_L(x,t))
 \end{align}
 where $\rho_{R}(x,t)$ and $\rho_{L}(x,t)$ are the (normal-ordered) right and left moving particle densities respectively, whereas the quantity in b) is,
   \begin{align}
\partial_t \Delta N(t)  \mbox{        } = \mbox{          }    \frac{d}{dt}  (N_R(t) - N_L(t))
 \end{align}
 where $ N_{R/L}(t) = \int dx \mbox{ } \rho_{R/L}(x,t) $. Naively, $   \Delta N(t)  \mbox{        } = \mbox{          }\int dx\mbox{ } \Delta \rho(x,t) $
  \\ \mbox{ } \\
  The striking result is this: in the limit when the bias becomes time independent, the answer for a) is zero but the answer for b) is non-zero. This is what we mean by bias-induced anomaly. Normally we expect that both should be zero since b) is obtained by spatially integrating a). The reason for this anomaly is that integrating over an infinite domain of x-values effectively multiplies the quantity by this infinity. Thus even if this quantity was tending to zero, when multiplied by a quantity tending to infinity, leads to a value that could be and in this case, is – finite.
\\ \mbox{ } \\
To see this mathematically we write,
 \begin{align}
 \partial_t \Delta \rho(x,t)  \mbox{        } &= \mbox{          }  \lim_{x^{'} \rightarrow x } v_F\frac{d}{dt}\bigg(<\psi^{\dagger}_{R}(x^{'},t)\psi_{R}(x,t) > -<\psi^{\dagger}_{L}(x^{'},t)\psi_{L}(x,t)>\bigg) \nonumber \\ \mbox{        } &= \mbox{          }-  \frac{v_F}{\pi}
\mbox{          }
    \left( \frac{\Gamma }{v_F} \mbox{          } \frac{(2 v_F)^2
}{\Gamma ^2 +4 v_F^2}\right)^2 \mbox{          }
   \frac{1}{2v_F}   \mbox{    } e V^{'}_b(t - \frac{|x|}{v_F})
 \end{align}
 whereas,
  \begin{align}
 \partial_t \Delta N(t)    \mbox{        } &= \mbox{          } - \frac{v_F}{\pi}
\mbox{          }
    \left( \frac{\Gamma }{v_F} \mbox{          } \frac{(2 v_F)^2
}{\Gamma ^2 +4 v_F^2}\right)^2 \mbox{          }
  \mbox{    } e   \mbox{ } [ V_b(t) - V_b(-\infty) ]
 \end{align}
Select $ V_b(t) = e^{ \alpha t } V_b(0) $ with $ \alpha \rightarrow 0^{+} $. This means the bias is slowly switched on from a zero value in the remote past and remains turned on at least until time $ t $. In the limit $ \alpha \rightarrow 0^{+} $, the bias is always on but the endpoints are mathematically well defined. In this case,
 \begin{align}
 \partial_t \Delta \rho(x,t)   \mbox{        } &= \mbox{          }- Lim_{\alpha \rightarrow 0^{+} } \frac{v_F}{\pi}
\mbox{          }
    \left( \frac{\Gamma }{v_F} \mbox{          } \frac{(2 v_F)^2
}{\Gamma ^2 +4 v_F^2}\right)^2 \mbox{          }
   \frac{1}{2v_F}   \mbox{    } e V^{'}_b(t - \frac{|x|}{v_F}) \approx 0
 \end{align}
 whereas,
  \begin{align}
 \partial_t \Delta N(t)    \mbox{        } &= \mbox{          } - Lim_{\alpha \rightarrow 0^{+} } \frac{v_F}{\pi}
\mbox{          }
    \left( \frac{\Gamma }{v_F} \mbox{          } \frac{(2 v_F)^2
}{\Gamma ^2 +4 v_F^2}\right)^2 \mbox{          }
  \mbox{    } e   \mbox{ } [ V_b(t) - V_b(-\infty) ] \nonumber \\&= \mbox{ } - Lim_{\alpha \rightarrow 0^{+} } \frac{v_F}{\pi}
\mbox{          }
    \left( \frac{\Gamma }{v_F} \mbox{          } \frac{(2 v_F)^2
}{\Gamma ^2 +4 v_F^2}\right)^2 \mbox{          }
  \mbox{    } e   \mbox{ }   V_b(0)  \neq 0
 \end{align}
   This is the first of the ``bias-induced anomalies".
\\ \mbox{ } \\ \mbox{ } \\
 The second example is the one-particle Green function itself - specifically the one which involves turning a right mover to a left mover i.e.
   $ < \psi^{\dagger}_L(x,t) \psi_R(x^{'},t^{'})> $. In the absence of a bias, this quantity vanishes identically when $ x $ and $ x^{'} $ are on opposite sides of the origin. However in the present case we obtain something interesting,
    \begin{align}
 <\psi^{\dagger}_{L}(x^{'} < 0,t^{'})\psi_{R}(x > 0,t)> = -\frac{i}{2\pi} \frac{\frac{\pi}{\beta v_{F}} \mbox{ }  q_1}{\sinh( \frac{ \pi }{\beta v_F } ( x + x^{'}-v_F(t-t^{'}) ) )} \mbox{     } ( -U(t - \frac{x}{v_F},t)    +
 U(t^{'} + \frac{x^{'}}{v_F},t)
   )    \end{align}
where $ q_1 =   \left[ 1
-       \frac{  2 \Gamma^2
}{\Gamma ^2 +4 v_F^2}
\right]
\mbox{ } i  \frac{\Gamma }{v_F}\frac{(2 v_F)^2
}{\Gamma ^2 +4 v_F^2}  $. This quantity is identically zero when there is no bias since the point contact causes reflection and turns a right mover to a left mover on the same side of the origin but not on opposite sides. But the presence of a bias  leads to a non-equilibrium situation where what is opposite sides of the origin at one time is effectively the same side of the origin at other times as seen by the presence of the evolution factor $ ( -U(t - \frac{x}{v_F},t)    +
 U(t^{'} + \frac{x^{'}}{v_F},t)
   )  $.

 \section{Dynamical density of states}
 \label{sec4}
 In this section, we present the formula for the dynamical density of states
 $D(\omega;x,T)$ at position $x$. $D(\omega;x,T) \mbox{    } d\omega$ is the number of fermionic states per unit length with energy between $\hbar \omega$ and $\hbar(\omega + d\omega)$ relative to the Fermi energy.  The DDOS at $x^{'}=x$ is given by the formula
 \begin{align}
 D(\omega;x,T)  \mbox{ } = \mbox{ }\int d\tau \mbox{    } e^{ -i \omega \tau } < \{ \psi(x,T + \frac{\tau }{2} ) , \psi^{\dagger}(x, T -  \frac{ \tau }{2} ) \}>
 \label{Domega}
 \end{align}
 A closed expression for $D(\omega;x,T)$ can be obtained by using the formulas for the non-equilibrium Green functions obtained earlier. We evaluate the DDOS for the right movers in \hyperref[AppendixB]{Appendix B} and it can be evaluated in a similar manner for the left movers as well. The DDOS for the right movers is obtained as
 \begin{align}
 D(\omega;x,T) \mbox{         }= \mbox{          }
D_{equil}(\omega;x,T) =  \frac{1}{2v_F}
 \end{align}
 In other words, it is the same as what one would expect when there is no bias, no mutual interactions between particles and the system is in equilibrium. The reason for this simple result is that in absence of mutual interactions between fermions, the anticommutator in eq.(\ref{Domega}) is a Dirac delta-function at $ \tau = 0 $ (since the space dependence and time dependence appear additively for chiral fermions).

\section{Time-dependent tunneling parameter}
\label{sec6}
In this section we consider the case of a time-dependent tunneling parameter $t_p$, which means $\Gamma$ is now time-dependent $\Gamma(t)$. We show that when the time-dependence of $\Gamma$ is present only as a phase factor $\Gamma_{TD}(t) = \Gamma e^{i \theta_{TD}(t)}$ the results are same as in the previous sections. But when the magnitude of $\Gamma$ itself is time-dependent i.e. $\Gamma_{TD}(t) = \Gamma(t)e^{i \theta_{TD}(t)}$ then these two approaches are not equivalent and we get different results. In general we can write a modified Hamiltonian for the system with a complex $\Gamma_{TD}(t)$,
\begin{align}
H = \sum_{p} v_Fp \mbox{ } b^{\dagger}_{p,R}b_{p,R} + \sum_{p}(-v_Fp) b^{\dagger}_{p,L}b_{p,L}+ \frac{ \Gamma_{TD}(t) }{L} b^{\dagger}_{.,R}b_{.,L} +\frac{ \Gamma^*_{TD}(t) }{L}b^{\dagger}_{.,L}b_{.,R}
 \label{timedepham}
\end{align}
Here we have used the same notations as in section \ref{sec1} and the Fermi fields are modified with the addition of a time-dependent phase $b_{p,R}(t) \mbox{     } = \mbox{      }c_{p,R}(t)\mbox{      }e^{ i \theta_{TD}(t) }$ and $b_{p,L}(t) \mbox{     } = \mbox{      }c_{p,L}(t) $. The equations of motion for the modified fields then become,
\begin{align}
i\partial_t b_{p,R}(t) \mbox{     } = \mbox{      } v_Fp  \mbox{      }  b_{p,R}(t) + \frac{ \Gamma_{TD}(t) }{L} b_{.,L}(t)\nonumber \\
i\partial_t b_{p,L}(t) \mbox{     } = \mbox{      } -v_Fp  \mbox{      } b_{p,L}(t) + \frac{  \Gamma^{*}_{TD}(t)  }{L} b_{.,R}(t)
\label{eqomb}
\end{align}

\subsection{ \texorpdfstring{$ |\Gamma_{TD}(t)| $}{Lg} is independent of time}
When the magnitude of the tunneling amplitude $|\Gamma_{TD}(t)|$ is independent of time we can write
\begin{align}
\Gamma_{TD}(t) = \Gamma \mbox{      }e^{ i \theta_{TD}(t) }
\end{align}
On comparing Eqs.\ref{eqomb} and \ref{eqom} so that,
\begin{align}
i\partial_t c_{p,R}(t)   \mbox{     } &= \mbox{      }( v_Fp
  + \partial_t\theta_{TD}(t) )   \mbox{      }  c_{p,R}(t)  + \frac{ |\Gamma|  }{L}c_{p,L}(t) \nonumber \\
  i\partial_t c_{p,L}(t) \mbox{     } &= \mbox{      } -v_Fp  \mbox{      }c_{p,L}(t) + \frac{ |\Gamma| }{L}c_{p,R}(t)
\end{align}
we can see that the case of $|\Gamma_{TD}(t)|$ independent of time is equivalent to the case of a time-independent real $\Gamma$ considered in the previous sections, provided one identifies $\partial_{t} \theta_{TD}(t) = e V_{b}(t)$. However when the time-dependence of the tunneling amplitude is such that the magnitude $|\Gamma_{TD}(t)|$ is dependent on time the two approaches are not equivalent.
\subsection{$|\Gamma_{TD}(t)|$ is time-dependent}
When the magnitude of the tunneling amplitude is itself time-dependent we may write,
\begin{align}
\Gamma_{TD}(t) = \Gamma(t) \mbox{      }e^{ i \theta_{TD}(t) }
\end{align}
Following a similar procedure as in Section \ref{sec1} we write down the finite temperature non-equilibrium Green functions for the case of a time-dependent tunneling parameter,
\begin{align}
 <\psi^{\dagger}_{\nu^{'}}(x^{'},t^{'})\psi_{\nu}(x,t)> \mbox{ }=\mbox{ } -\frac{i}{2\pi} \frac{\frac{\pi}{\beta v_{F}}}{\sinh( \frac{ \pi }{\beta v_F } (\nu x-\nu^{'}x^{'}-v_F(t-t^{'}) ) )} \zeta_{\nu,\nu^{'}}
\label{timedepnegf}
 \end{align}
 where $\nu$,$\nu^{'} = \pm 1$ with $R=1$ and $L=-1$ and
 \begin{widetext}
 \begin{align}
 \zeta_{1,1} = \left( 1 +  \frac{ i }{2v_F}      \mbox{   }
 \Gamma^*_{TD}(t-  \frac{x}{v_F})
\mbox{   } \Xi_R(x,t) \right)
\left( 1 - \frac{ i }{2v_F}      \mbox{   }
 \Gamma_{TD}(t^{'} -  \frac{x^{'}}{v_F})
\mbox{   } \Xi^*_R(x^{'},t^{'}) \right)
 +  \Xi_R(x,t)
 \mbox{  } \Xi^*_R(x^{'},t^{'})
 \end{align}
 \begin{align}
 \zeta_{1,-1} = \Xi^*_L(x^{'},t^{'})\mbox{      }\left( 1 +  \frac{ i }{2v_F}      \mbox{   }
 \Gamma^*_{TD}(t-  \frac{x}{v_F})
\mbox{   } \Xi_R(x,t) \right)-\Xi_R(x,t)
\mbox{      } \left( 1
 + \frac{i}{2v_F}  \mbox{      }
    \Gamma^*_{TD}(t^{'} + \frac{x^{'}}{v_F})
 \mbox{ } \Xi^*_L(x^{'},t^{'})\right)
 \end{align}
 \begin{align}
 \zeta_{-1,1} = \Xi_L(x,t)
\left( 1 - \frac{ i }{2v_F}      \mbox{   }
 \Gamma_{TD}(t^{'} -  \frac{x^{'}}{v_F})
\mbox{   } \Xi^*_R(x^{'},t^{'}) \right)-\Xi^*_R(x^{'},t^{'})\mbox{      } \left( 1
 - \frac{i}{2v_F}  \mbox{      }
    \Gamma_{TD}(t + \frac{x}{v_F})
 \mbox{ } \Xi_L(x,t)\right)
 \end{align}
 \begin{align}
 \zeta_{-1,-1} = \left( 1
 - \frac{i}{2v_F}  \mbox{      }
    \Gamma_{TD}(t + \frac{x}{v_F})
 \mbox{ } \Xi_L(x,t)\right)
\mbox{      } \left( 1
 + \frac{i}{2v_F}  \mbox{      }
    \Gamma^*_{TD}(t^{'} + \frac{x^{'}}{v_F})
 \mbox{ } \Xi^*_L(x^{'},t^{'})\right)
 +  \Xi_L(x,t)
 \mbox{      } \Xi^*_L(x^{'},t^{'})
 \end{align}
 \end{widetext}
where
\begin{align}
\Xi_R(x,t)\mbox{     } = \mbox{      } \frac{i}{v_F}  \mbox{      }
     \frac{  \theta(x  )  \mbox{   } \Gamma_{TD}(t-  \frac{x}{v_F})
 }{ 1 +  \frac{ |\Gamma_{TD}(t-  \frac{x}{v_F} )|^2 }{(2v_F)^2 }
  }; \mbox{ } \mbox{  }
  \Xi_L(x,t)\mbox{     } = \mbox{      } - \frac{ i }{v_F}
 \frac{        \theta( -x  )   \mbox{   } \Gamma^*_{TD}(t + \frac{ x }{ v_F} ) }{1 +  \frac{ |\Gamma_{TD}(t +  \frac{x}{v_F} )|^2 }{(2v_F)^2 }}
 \label{xiequil}
\end{align}
 The tunneling current can be computed using the same definition as in Eq.\ref{tundef} but with the Hamiltonian given in Eq.\ref{timedepham}. This gives,
 \begin{align}
  I_{tun}(t) \mbox{        } = \mbox{          }-i \mbox{ }e \lim_{t^{'} \rightarrow t } \bigg(
  \Gamma_{TD}(t) \mbox{ } < \psi^{\dagger}_R(0,t^{'})  \psi_L(0,t) >  -  \Gamma^*_{TD}(t) \mbox{ } < \psi^{\dagger}_L(0,t)  \psi_R(0,t^{'}) > \bigg)
  \label{ituntimedep}
 \end{align}
After some effort we conclude,
 \begin{align}
 I_{tun}(t) \mbox{        } = \mbox{          }\frac{4 i e v_F^2 \left( \Gamma_{ TD }^*(t) \Gamma_{TD}^{'}(t)-
 \Gamma_{TD}(t) \Gamma_{TD}^{*'}(t)\right)}{\pi  \left( | \Gamma_{TD}(t) |^2+4 v_F^2\right)^2}
 \label{itungammat}
 \end{align}
 As the magnitude of $\Gamma_{TD}(t)$ is time-dependent in this case we set $ \Gamma_{TD}(t) \mbox{        } = \mbox{          } \Gamma (t)  \mbox{          }e^{i  \theta_{TD}(t)} $ and using the relation $ \theta_{TD}'(t) = e V_b(t) $ we get
 \begin{align}
 I_{tun}(t) \mbox{        } = \mbox{          }-\frac{8 e^2 v_{F}^2 \Gamma (t)^2  }{\pi  \left(\Gamma (t)^2+4 v_{F}^2\right)^2}\mbox{ }\mbox{ } V_b(t)
 \end{align}
 The time-dependent tunneling amplitude in terms of the tunneling parameter $t_{p}(t)$ can be written as $\Gamma(t) = 2 v_{F} t_{p}(t)$. Also using the convention as in \cite{shah2016consistent} we pick up a minus sign $V_{b}(t) = -V(t)$. Finally the following expression for the time-dependent tunneling current is obtained after restoring dimensional units
 \begin{align}
 I_{tun}(t) \mbox{        } = \mbox{          }\frac{ e^2   }{h } \mbox{ } \frac{ 4\mbox{ } t_p (t)^2   }{
 \left(1+t_p(t)^2\right)^2}\mbox{ }V(t)
 \end{align}
  The above expression is similar in form to Eq. \ref{eqtunc} except that $t_{p}$ is now time-dependent and this tells us that so long as $ t_p(t) $ is switched on and held at a constant value, the current is proportional to voltage and there are no transients in the current. However if $ t_p(t) $ itself depends in some complicated way on the bias, this may not be the case. In such situations, this dependence has to be specified separately.
 \section{Double barrier resonant tunneling}
 \label{db}
\begin{figure}[H]
\centering
 \includegraphics[scale=0.9]{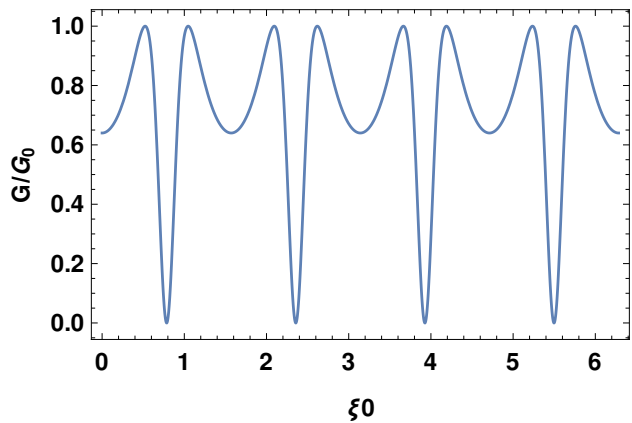}
 \caption{This figure shows $G/G_{0}$ vs $\xi_{0}$ for a symmetric double barrier. The conductance peaks when the resonance condition (Eq.\ref{rescondsym}) is satisfied. The figure is plotted choosing $W=2$ and $v_{F}=1$.}
 \label{resonant}
 \end{figure}

 The type of impurity present at the point contact is given by the form of $\Gamma_{TD}(t)$. Here we show that the results from the previous section can be used to easily study resonant tunneling by treating the point contact impurity as a double barrier structure such as a lead-device-lead junction at the origin. We neglect Coulomb interactions and dynamics in the device and only focus on point-contact coupling between the left-lead and the device and the right-lead and the device. In this case the tunneling amplitude takes the form
 \begin{align}
 \Gamma_{TD}(t) = \Gamma(t) e^{i \Theta_{TD}} = (W_{L} e^{2 i \xi_{0}} + W_{R} e^{-2 i \xi_{0}})e^{i \Theta_{TD}(t)}
 \label{dbgamma}
 \end{align}
 where $\Theta^{'}_{TD}(t) = e V_{b}(t) = -e V(t)$ and the coupling between the double barrier at the origin and the left and right chiral quantum wires is $W_{L}$ and $W_{R}$ respectively which can be time-dependent in general but we assume them to be independent of time for simplicity as it doesn't change the form of the I-V characteristics. Eq.\ref{dbgamma} is valid in the limit $k_{F} \rightarrow \infty$ and inter-barrier separation $a \rightarrow 0$ such that $0 < k_{F} a = \xi_{0} < \infty$ is fixed. For the case of a symmetric double barrier $W_{L} = W_{R} = W$ and substituting Eq.\ref{dbgamma} in Eq.\ref{itungammat} we obtain
 \begin{align}
 I_{tun}(t) = \frac{8 e^{2} v_{F}^{2} W^{2}(e^{2 i \xi_{0}} + e^{-2 i \xi_{0}})^{2}}{\pi(4 v_{F}^{2} + W^{2}(e^{2 i \xi_{0}} + e^{-2 i \xi_{0}})^{2})^{2}} V(t)
 \end{align}
 The tunneling conductance is given by
 \begin{align}
 G = \frac{8 e^{2} v_{F}^{2} W^{2}(e^{2 i \xi_{0}} + e^{-2 i \xi_{0}})^{2}}{\pi(4 v_{F}^{2} + W^{2}(e^{2 i \xi_{0}} + e^{-2 i \xi_{0}})^{2})^{2}}
 \end{align}
 This is of the form $G = \frac{8 e^{2} v_{F}^{2} |\Gamma|^{2}}{\pi(4 v_{F}^{2} + |\Gamma|^{2})^{2}}$, which for $|\Gamma|^{2} = 4 v_{F}^{2}$ peaks at $G = \frac{e^{2}}{2\pi} = \frac{e^{2}}{h}$ (since $\hbar = 1$). This is used to determine the condition for resonant tunneling.
 One can tune the inter-barrier separation in the device region by external means in order to achieve resonance. When the resonance condition is satisfied the conductance attains its peak value $G = G_{0} = \frac{e^{2}}{h}$ (see Fig.\ref{resonant}).
 The condition for resonance in the symmetric double barrier case is determined to be
 \begin{align}
 W^{2}(2 \cos(4 \xi_{0}) + 2)-4 v_{F}^{2} = 0
 \label{rescondsym}
 \end{align}
 . For the more general case of an asymmetrical double barrier the expression for current is obtained as
 \begin{align}
 I_{tun}(t) = \frac{8 e^{2} v_{F}^{2}\left(e^{2 i \xi_{0}} W_{L}+e^{-2 i \xi_{0}} W_{R}\right) \left(e^{-2 i \xi_{0}} W_{L}+e^{2 i \xi_{0}} W_{R}\right)}{\pi (4 v_{F}^{2} + \left(e^{2 i \xi_{0}} W_{L}+e^{-2 i \xi_{0}} W_{R}\right) \left(e^{-2 i \xi_{0}} W_{L}+e^{2 i \xi_{0}} W_{R}\right))^{2}}V(t)
 \end{align}
 The tunneling conductance attains its maximum value $G_{0} = \frac{e^{2}}{h}$ when the resonance condition for this case is satisfied, which is
 \begin{align}
 W_{L}^{2} + 2 \cos(4 \xi_{0}) W_{L} W_{R} + W_{R}^{2} - 4 v_{F}^{2} = 0
 \end{align}
 \section{Transient quantum transport in the case of finite bandwidth}
 \label{secfinite}
 So far what we have considered is the simplest possible scenario of non-interacting fermions with a linear dispersion and no momentum cutoffs (infinite bandwidth). In such a circumstance the I-V characteristic (Eq.\ref{eqtunc})  for point-contact tunneling is linear and only shows steady state behaviour when a constant voltage is turned on suddenly. We expect deviations from these properties when we take into account the effects of a momentum cutoff (finite bandwidth) $\Lambda$ in the point contact such that $|p| < \Lambda$.  This implicitly means that there is a short distance cutoff ($\frac{1}{\Lambda}$) in the problem, which makes the quantum point contact (QPC) region to occupy a non-zero size. Here we calculate the leading correction to the tunneling current due to a finite bandwidth and show the appearance of a transient in the current in presence of a (subsequently constant) bias that is suddenly switched on. The correction to the current involves integration over the past history of the system and thus encodes memory effects ie. it displays non-Markovian behaviour.\\ \mbox{ } \\
 In presence of a finite bandwidth, the Hamiltonian is written as
 \begin{align}
 H = \sum_p(v_F p + eV_b(t))c^{\dagger}_{p,R}c_{p,R} + \sum_p(-v_F p)c^{\dagger}_{p,L}c_{p,L}
 + \frac{\Gamma}{L}(c^{\dagger}_{.,R}c_{.,L} + c^{\dagger}_{.,L}c_{.,R})
 \label{hamfinite}
 \end{align}
 where now we have defined $ c_{.,\nu} = \sum_{|p|<\Lambda}c_{p,\nu} $ with the momentum restricted to a finite bandwidth $\Lambda$. We set $ c_{.,\nu} = c^{\infty}_{.,\nu} + \delta c_{.,\nu} $ where $ c^{\infty}_{.,\nu} = \sum_pc_{p,\nu} $ and $\delta c_{.,\nu}$ is the deviation from the infinite bandwidth case. The other quantities have their usual meaning as in the previous sections. The equations of motion for the right and left moving Fermi fields in this case are
 \begin{align}
 i \partial_t c_{p,R}(t) = (v_Fp + eV_b(t)) c_{p,R}(t) + \theta(\Lambda-|p|)\frac{\Gamma}{L}c_{.,L}(t) \nonumber \\
 i \partial_t c_{p,L}(t) = -v_Fp c_{p,L}(t) + \theta(\Lambda-|p|)\frac{\Gamma}{L}c_{.,R}(t)
 \label{bandeqom}
 \end{align}
 with $\theta(\Lambda-|p|)$ taken to be the Dirichlet regularized step function. Exact solution to the problem is possible only in the simplest case of infinite bandwidth. If we wish to investigate the case of a finite bandwidth analytically, we shall have to settle for perturbative corrections in powers of $\frac{1}{\Lambda}$. In this section, we are interested in the finite bandwidth correction to the current $\delta I_{tun}(t)$ and we shall evaluate it upto $O(\frac{1}{\Lambda})$ i.e. we assume a large but finite bandwidth. In this case we may write,
 \begin{align}
 \delta I_{tun}(t) =& - \frac{ i e \Gamma}{L} Lim_{t^{'} \rightarrow t }(<\delta c^{\dagger}_{.,R}(t^{'})c^{\infty}_{.,L}(t)>
- <\delta c^{\dagger}_{.,L}(t)c^{\infty}_{.,R}(t^{'})>)\nonumber \\&
 - \frac{ i e \Gamma}{L} Lim_{t^{'} \rightarrow t }(<c^{\dagger \infty}_{.,R}(t^{'})\delta c_{.,L}(t)>- <c^{\dagger \infty }_{.,L}(t)\delta c_{.,R}(t^{'})>)
 \label{ditun}
 \end{align}
 Upon solving the equations in Eq.\ref{bandeqom} we may write,
 \begin{align}
 c_{.,R}(t) = \left(  \sum_{|p|<\Lambda} c_{p,R}(t_0) e^{ -i (t-t_0)  p v_F  }\right)\mbox{      }  U(t_0,t)- i \Gamma  \int_{t_0}^t c_{.,L}(t_2)
  \delta_{\Lambda} (v_F(t-t_2))\mbox{ } U(t_2,t) \, dt_2
  \label{cdotR}
 \end{align}
 and
 \begin{align}
  c_{.,L}(t) =  \left(  \sum_{|p|<\Lambda} c_{p,L}(t_0) e^{ i (t-t_0)  p v_F  }\right)- i \Gamma   \int_{t_0}^t c_{.,R}(t_2)
 \delta_{\Lambda}(v_F(t-t_2))  \, dt_2
 \label{cdotL}
 \end{align}
 where we have defined the broadened Dirac delta as $\delta_{\Lambda}(x) = \frac{1}{L}\sum_{|p|<\Lambda}e^{ipx}$. We find the need to introduce a quantity $ \Delta(x) $ that corresponds to the difference between this broadened and actual Dirac delta function. In terms of the usual Dirac delta function $\delta(X) = \frac{1}{L} \sum_{p} e^{i p X}$ this can be written as
 \begin{align}
 \delta_{\Lambda} (X) = \delta(X) - \Delta(X)
 \label{deltalambda}
 \end{align}
 Computing the finite bandwidth correction to leading order in $ 1/\Lambda $ is itself not a straightforward task and involves a tedious calculation. In order to calculate $\delta I_{tun}(t)$ we have to evaluate correlations of the type $<\delta c^{\dagger}_{.,\nu^{'}}(t^{'})c^{\infty}_{.,\nu}(t)>$ and $<c^{\dagger \infty}_{.,\nu^{'}}(t^{'})\delta c_{.,\nu}(t)>$. In \hyperref[AppendixD]{Appendix D} we discuss the procedure to calculate these correlations. After a lengthy procedure we obtain the following expression
 \begin{widetext}
 \begin{align}
\delta I_{tun}(t) = &-\frac{i e \Gamma}{L}\bigg( -\frac{4 L v_F^2 \left(4 v_F^2-\Gamma^2\right) (U(t,t_0)-U(t_0,t))}{ (\Gamma ^2+4 v_F^2)^2}  \int dx
    \frac{\Delta(x-v_{F}(t-t_{0}))}{2\pi i} \frac{ \frac{ \pi }{\beta v_F } \mbox{ }\theta(-x)  (R-R^*) }{\sinh( \frac{ \pi }{\beta v_F } (-x + v_{F}(t-t_{0}) ) )  }
\mbox{              }
    \mbox{          } \nonumber \\
    &+ \frac{32 i \Gamma^3 L v_F^4}{(\Gamma^2+4 v_F^2)^3}
 \mbox{   }  \int_{t_0}^{t}  \Delta(v_F(t-t_2))\mbox{ }
    \frac{1}{2\pi i} \frac{ \frac{ \pi }{\beta v_F } }{\sinh( \frac{ \pi }{\beta} (t-t_{2})  )  }
\mbox{          }
  (U(t,t_2) U(t,t_2)- U(t_2,t)  \mbox{ } U(t_2,t) ) \, dt_2 \bigg)
  \label{deltaifin}
 \end{align}
 \end{widetext}
 \begin{figure}
\centering
 \includegraphics[scale=1]{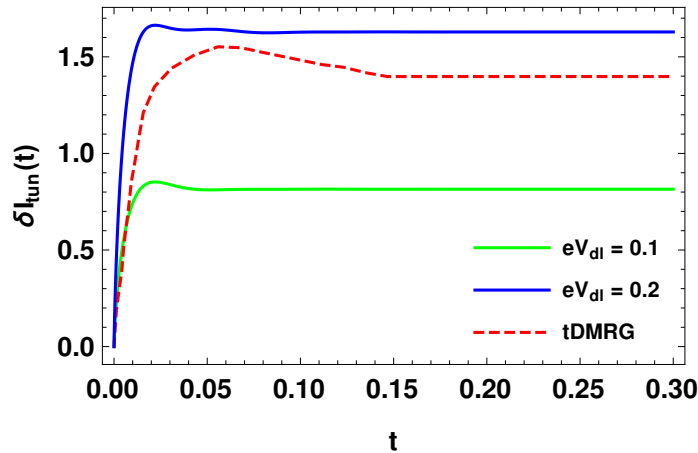}
 \caption{Current transients: \small This figure shows $\delta I_{tun}(t)$ vs $t$ for different values of dimensionless bias $e V_{dl}$ defined as $e V_{dl} = \frac{e V_{0}}{v_{F} \Lambda}$ where we have considered a step bias $V(t) = \theta(t) V_{0}$ and for dimensionless temperature $T_{dl} = \frac{T}{v_{F} \Lambda} = 0.1$. The other parameters are $\Gamma \Lambda = 100$ and $v_{F} \Lambda = 100 $ in appropriate units. The red-dashes indicate the $tDMRG$ result for the time-evolution of current in an Anderson dot model (see Fig.1 in \cite{Eckel_2010}), which we have rescaled and overlayed on our plot to show the qualitative similarity between the current transients even in a completely different model to ours.}
 \label{deltaivst}
 \end{figure}
 Here $R$ is the reflection amplitude (for the infinite band-width point contact), $\beta$ is inverse temperature, $x$ is position and $t_{0}$ is an arbitrary initial time when the system is in equilibrium. Since we are going to set $t_{0} \rightarrow -\infty$ as we did in the case of infinite bandwidth, the first term in Eq.\ref{deltaifin} drops out and we are left with only the second term which involves an integral over the past history of the system.
 \begin{widetext}
 \begin{align}
 \delta I_{tun}(t) = -\frac{i e \Gamma}{L} \frac{32 i \Gamma^3 L v_F^4}{(\Gamma^2+4 v_F^2)^3}
 \mbox{   }  \int_{-\infty}^{t}
    \frac{\Delta(v_F(t-t_2))}{2\pi i} \frac{ \frac{ \pi }{\beta v_F } }{\sinh( \frac{ \pi }{\beta} (t-t_{2})  )  }
  (U(t,t_2) U(t,t_2)- U(t_2,t)  \mbox{ } U(t_2,t) ) \, dt_2
  \label{deltaifinal}
 \end{align}
 \end{widetext}
Note that the function $\Delta $ may be written as $\Delta(v_F(t-t_2)) = \delta(v_F(t-t_2)) - \frac{\sin(\Lambda v_F(t-t_2))}{\pi v_F(t-t_2)} $. The expression for $\delta I_{tun}(t)$ in Eq.\ref{deltaifinal} exhibits non-Markovian behaviour as the current at a given time is depends not only on the voltage applied at that time but also on the voltage at all previous times.
 When the bias is switched on suddenly and remains constant thereafter, the tunneling current shows a transient (gradual) build up before settling into its steady state constant value. Even when a constant voltage is present eternally, there is a non-linear dependence of the current on this bias voltage which is seen in terms such as $U(t,t_2)$, where $U(t,t_{2}) \equiv e^{ - i \int _{ t}^{t_{2}}  e V_{b}(s)ds }$. Note that according to our convention $V_{b}(t) = -V(t)$ and hence $U(t,t_{2}) \equiv e^{  i \int _{ t}^{t_{2}}  e V(s)ds }$ (where $ V(t) $ is the convention of \cite{shah2016consistent}). In Fig.\ref{deltaivst} we plot $\delta I_{tun}(t)$ vs $t$ for the case of a step bias (sudden switching on) $V(t) = V_{0} \mbox{ }\theta(t)$ showing the transients in the current before reaching a steady state. 
 The transients are qualitatively similar to that observed in the experimental investigation of split-gate quantum point-contacts \cite{doi:10.1063/1.2337865} and also in numerical simulations of non-equilibrium transport through an Anderson dot using methods like time-dependent density matrix renormalization group ($tDMRG$) \cite{Eckel_2010} and the iterative summation of path integrals ($ISPI$) \cite{PhysRevB.77.195316} approach.\\
 Our analysis shows that even in non-equilibrium transport through a simple quantum point-contact, transient dynamics appear in the tunneling current when a short distance cutoff is introduced in the problem. This is the reason why numerical methods like $tDMRG$, that work on a lattice, predict a transient in the current before a steady state is reached. Fig.\ref{deltaivdltdl} shows the steady state current $\delta I_{tun}$ as a function of constant time-independent bias ($e V_{dl} = \frac{e V}{v_{F} \Lambda}$) for different temperatures $T_{dl} = \frac{T}{v_{F} \Lambda}$. The three energy scales in the problem are temperature ($T$), bias potential ($V$) and the energy scale associated with the bandwidth ($v_{F} \Lambda$). The energy scale due to the bandwidth is the dominant one as we assume the bandwidth to be large but finite in our calculation.
 \begin{figure}[H]
 \centering
 \includegraphics[scale=0.9]{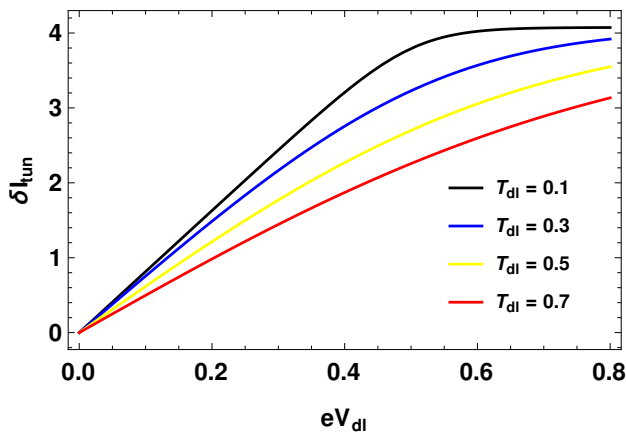}
 \caption{Nonlinear I-V characteristics: \small The temperature dependence of $\delta I_{tun}$ in steady state is shown when plotted vs $e V_{dl} = \frac{e V}{v_{F} \Lambda}$. The dimensionless temperature is defined as $ T_{dl} = \frac{T}{v_{F} \Lambda}$. The other parameters are $\Gamma \Lambda = 100 $ and $v_{F} \Lambda = 100$ in appropriate units.}
 \label{deltaivdltdl}
\end{figure}

\section{Summary and Conclusions}
\label{conc}
In this paper, we have investigated the non-equilibrium transport between two non-interacting chiral quantum wires by computing exactly the non-equilibrium Green function (NEGF) expressed in terms of simple functions of positions and times unlike previous analytical approaches that deal only with the steady state response to a constant bias. We have calculated the time-dependent tunneling current and differential conductance across an infinite bandwidth point contact for an arbitrary time-dependent bias. In this case no transients in the tunneling current are seen when a step voltage bias is considered $V_{b}(t) = \theta(t)\mbox{ }  V_{0}$. In this case, the steady state in the current is reached instantaneously which may be attributed to the extreme ideal situation of absence of interactions between fermions and infinite bandwidth in the point contact. The NEGF method allows us to study transient phenomena as well as steady state properties. The full space-time NEGF that we have obtained exhibits transient behavior upon sudden switch on of a bias even when the bandwidth of the point-contact is infinite. We also examine the situation where in addition to the bias voltage, the tunneling amplitude also becomes time-dependent. We calculate the NEGF and the time-dependent transport properties in such a scenario. We also demonstrate how resonant tunneling through a simple double barrier structure can be easily studied using our method.\\

 We go beyond the infinite bandwidth limit and consider the situation when a finite bandwidth ($\Lambda$) in the point-contact is introduced in the problem. Although exact treatment of time-dependent non-equilibrium transport in a similar system has been studied before using Keldysh NEGF formalism, our method is different as we work in the position and time domain and our method applies for an arbitrary time-dependent bias unlike previous works \cite{PhysRevB.74.085324} that are restricted to special cases like a sharp step bias or a square pulse voltage bias. When a finite bandwidth is introduced in the point-contact, a short distance cutoff ($\frac{1}{\Lambda}$) becomes implicit. A systematic perturbative treatment in this parameter allows the calculation of the correction to the tunneling current upto $O(\frac{1}{\Lambda})$. We have shown that the transport properties are non-Markovian in this case. The tunneling current now shows transient behaviour before reaching a steady state (for a bias that is suddenly switched on) which is merely a consequence of the presence of a short-distance cutoff in the problem description and not on other details.\\

 In addition to the non-equilibrium two-point function, it is also possible to write down (using Wick's theorem) the four-point functions. Using these correlations in conjunction with powerful novel bosonization techniques \cite{das2018quantum} it is possible to extend our NEGF approach to study non-equilibrium transport between chiral fermionic edges with mutually interacting particles like in the case of Fractional Quantum Hall edge states \cite{chang2003chiral}. This approach could shed more light on the universality of power-law exponents, or more generally, the scaling functions \cite{PhysRevB.52.8934}, in systems with chiral Luttinger liquid character.\\

\section*{Acknowledgements}

We gratefully acknowledge the reactions of the authors cited in this paper, specially C.J. Bolech who suggested that we investigate appearance of a transient in the current as predicted in earlier works using numerical methods like $tDMRG$.


\section*{APPENDIX A: Calculation of tunneling current}
\label{AppendixA}
\setcounter{equation}{0}
\renewcommand{\theequation}{A.\arabic{equation}}
The tunneling current is
\begin{align}
I_{tun}(t) \mbox{        } = \mbox{          }-i e \Gamma \mbox{        } \lim_{t^{'} \rightarrow t } \bigg( < \psi^{\dagger}_R(0,t^{'})  \psi_L(0,t) > - < \psi^{\dagger}_L(0,t)  \psi_R(0,t^{'}) > \bigg)
\end{align}
From Eqs.\ref{eqnoneq}-\ref{eqkappa4} and using the Dirichlet criterion $\theta(0) = \frac{1}{2}$ we can write
\begin{align}
<\psi^{\dagger}_{R}&(0,t^{'})\psi_{L}(0,t)>   \mbox{        } = \mbox{          }
- \frac{i}{2\pi} \frac{ \frac{ \pi }{\beta v_F } }{\sinh( \frac{ \pi }{\beta v_F } (-v_F(t-t^{'}) ) )  }
 \left( - U(t^{'},t) + 1  \right)\mbox{ }i \frac{\Gamma }{v_F}    \frac{   2  v_F^2   }{\Gamma^2 +4 v_F^2} \left[1
-       \frac{    \Gamma^2
}{\Gamma ^2 +4 v_F^2}\right]
\end{align}
then we get,
\begin{align}
 I_{tun}(t) \mbox{        } =  \mbox{          }i e \Gamma \mbox{        }  \lim_{t^{'} \rightarrow t }\mbox{        }  \bigg(   \frac{1}{2\pi} \frac{ \frac{ \pi }{\beta v_F } }{\sinh( \frac{ \pi }{\beta v_F } (v_F(t-t^{'}) ) )  }
 \left(  U(t,t^{'}) - U(t^{'},t)  \right) \mbox{          }   \frac{\Gamma }{v_F}    \frac{   2  v_F^2   }{\Gamma ^2 +4 v_F^2} \left[1
-       \frac{    \Gamma^2
}{\Gamma ^2 +4 v_F^2}\right] \bigg)
\end{align}
Note that we have defined $U(t,t^{'}) \mbox{        } = \mbox{          } e^{ -i \int^{t^{'}}_{t} d\tau \mbox{    } e V_b(\tau) }$. Evaluating the limit using L'Hospital's rule,
\begin{align}
  I_{tun}(t) \mbox{        } = \mbox{          } -   \Gamma^2 \mbox{        }
     \mbox{          }    \frac{   4   }{\Gamma ^2 +4 v_F^2} \left[1
-       \frac{    \Gamma^2
}{\Gamma ^2 +4 v_F^2}\right]
\mbox{    }
\frac{e^2}{2\pi } V_b(t)
\end{align}
Expressing the tunneling amplitude in terms of a tunneling parameter $t_{p}$ we get the expression for the tunneling current as in the main text Eq.\ref{eqtunc}.
\section*{APPENDIX B: DDOS for right movers}
\label{AppendixB}
\setcounter{equation}{0}
\renewcommand{\theequation}{B.\arabic{equation}}
The dynamical density of states for the right movers is given by the equation
\begin{align}
D(\omega;x,T)  \mbox{ } = \mbox{ }\int d\tau \mbox{    } e^{ -i \omega \tau } < \{ \psi(x,T + \frac{\tau }{2} ) , \psi^{\dagger}(x, T -  \frac{ \tau }{2} ) \}>
\end{align}
Using Eq.\ref{eqnoneq} we can write,
\begin{align}
& <  \{ \psi_{R}(x,T + \frac{\tau }{2} ) , \psi^{\dagger}_{R}(x^{'},T-\frac{\tau }{2}) \}> \mbox{        }= \mbox{ }
   \frac{i}{2\pi} \bigg( \frac{ \frac{ \pi }{\beta v_F } }{\sinh( \frac{ \pi }{\beta v_F } (x-x^{'}-v_F \tau + i v_F \epsilon ) )  }
 - \frac{ \frac{ \pi }{\beta v_F } }{\sinh( \frac{ \pi }{\beta v_F } (x-x^{'}-v_F \tau - i v_F \epsilon ) )  }  \bigg)
 \nonumber \\ &  \mbox{          }
    \bigg(  U(T -  \frac{ \tau }{2} ,T + \frac{\tau }{2}) \left[1
-      \theta(x^{'})    \mbox{          }   \frac{  2 \Gamma^2
}{\Gamma ^2 +4 v_F^2}\right]\mbox{          }    \left[1
-      \theta(x)    \mbox{          }   \frac{  2 \Gamma^2
}{\Gamma ^2 +4 v_F^2}\right]\nonumber \\ & + \left( \frac{\Gamma }{v_F} \mbox{          } \frac{(2 v_F)^2
}{\Gamma ^2 +4 v_F^2}\right)^2 \mbox{          }
  \theta(x )     \theta(x^{'} )   \mbox{          }  U(T -  \frac{ \tau }{2} ,T -  \frac{ \tau }{2}  - \frac{x^{'}}{v_F})   \mbox{     }
   U(T + \frac{\tau }{2}  - \frac{x}{v_F},T + \frac{\tau }{2} )   \bigg)
\end{align}
where we have taken $ t = T + \frac{\tau }{2}   $ and  $ t^{'} = T -  \frac{ \tau }{2}  $ and $\epsilon > 0$. In the zero temperature limit $\beta \rightarrow \infty$ doing an expansion in powers of $\epsilon$ and finally taking the limit $\epsilon \rightarrow 0$ we can write
\begin{align}
 & \frac{i}{2\pi} \bigg(  \frac{ \frac{ \pi }{\beta v_F } }{\sinh( \frac{ \pi }{\beta v_F } (x-x^{'}-v_F \tau + i v_F \epsilon ) )  }
  - \frac{ \frac{ \pi }{\beta v_F } }{\sinh( \frac{ \pi }{\beta v_F } (x-x^{'}-v_F \tau - i v_F \epsilon ) )  }  \bigg)
 \mbox{        }= \mbox{          }\\ & \frac{\epsilon/\pi }{2  v_F \left((\tau  + \frac{ -x+x^{'} }{v_F} )^2+\epsilon ^2\right)}= \frac{1}{2v_F}
 \delta(\tau  + \frac{ -x+x^{'} }{v_F} )
\end{align}
This means,
\begin{align}
&< \{ \psi_{R}(x,T + \frac{\tau }{2} ) , \psi^{\dagger}_{R}(x^{'}, T -  \frac{ \tau }{2} ) \}> \mbox{        } \nonumber \\ =& \mbox{          }
\frac{1}{2v_F}
 \delta(\tau  + \frac{ -x+x^{'} }{v_F} )U(T -  \frac{ x-x^{'} }{2v_F} ,T + \frac{ x-x^{'} }{2v_F})\mbox{          }
\mbox{          }    \bigg(  \left[1
-      \theta(x^{'})    \mbox{          }   \frac{  2 \Gamma^2
}{\Gamma ^2 +4 v_F^2}\right]\mbox{          }    \left[1
-      \theta(x)    \mbox{          }   \frac{  2 \Gamma^2
}{\Gamma ^2 +4 v_F^2}\right]
\nonumber \\ & + \left( \frac{\Gamma }{v_F} \mbox{          } \frac{(2 v_F)^2
}{\Gamma ^2 +4 v_F^2}\right)^2 \mbox{          }
  \theta(x )     \theta(x^{'} )    \bigg)
\end{align}
At $x^{'} = x$, the DDOS for the right movers is,
\begin{align}
& D(\omega;x,T)  \mbox{        }= \mbox{          }\int d\tau \mbox{    } e^{ -i \omega \tau } < \{ \psi_{R}(x,T + \frac{\tau }{2} ) , \psi^{\dagger}_{R}(x, T -  \frac{ \tau }{2} ) \}> \mbox{        }\nonumber \\ =& \mbox{          }
\frac{1}{2v_F}
  \mbox{          }
    \bigg(  \left[1
-      \theta(x)    \mbox{          }   \frac{  2 \Gamma^2
}{\Gamma ^2 +4 v_F^2}\right]^2
 + \left( \frac{\Gamma }{v_F} \mbox{          } \frac{(2 v_F)^2
}{\Gamma ^2 +4 v_F^2}\right)^2 \mbox{          }
  \theta(x )        \bigg)\mbox{ } = \mbox{ }\frac{1}{2v_F}
\end{align}

 \section*{APPENDIX C: Relation to the Keldysh non-equilibrium Green function}
\label{AppendixC}
\begin{figure}[H]
\subfigure[]{\includegraphics[scale=0.4]{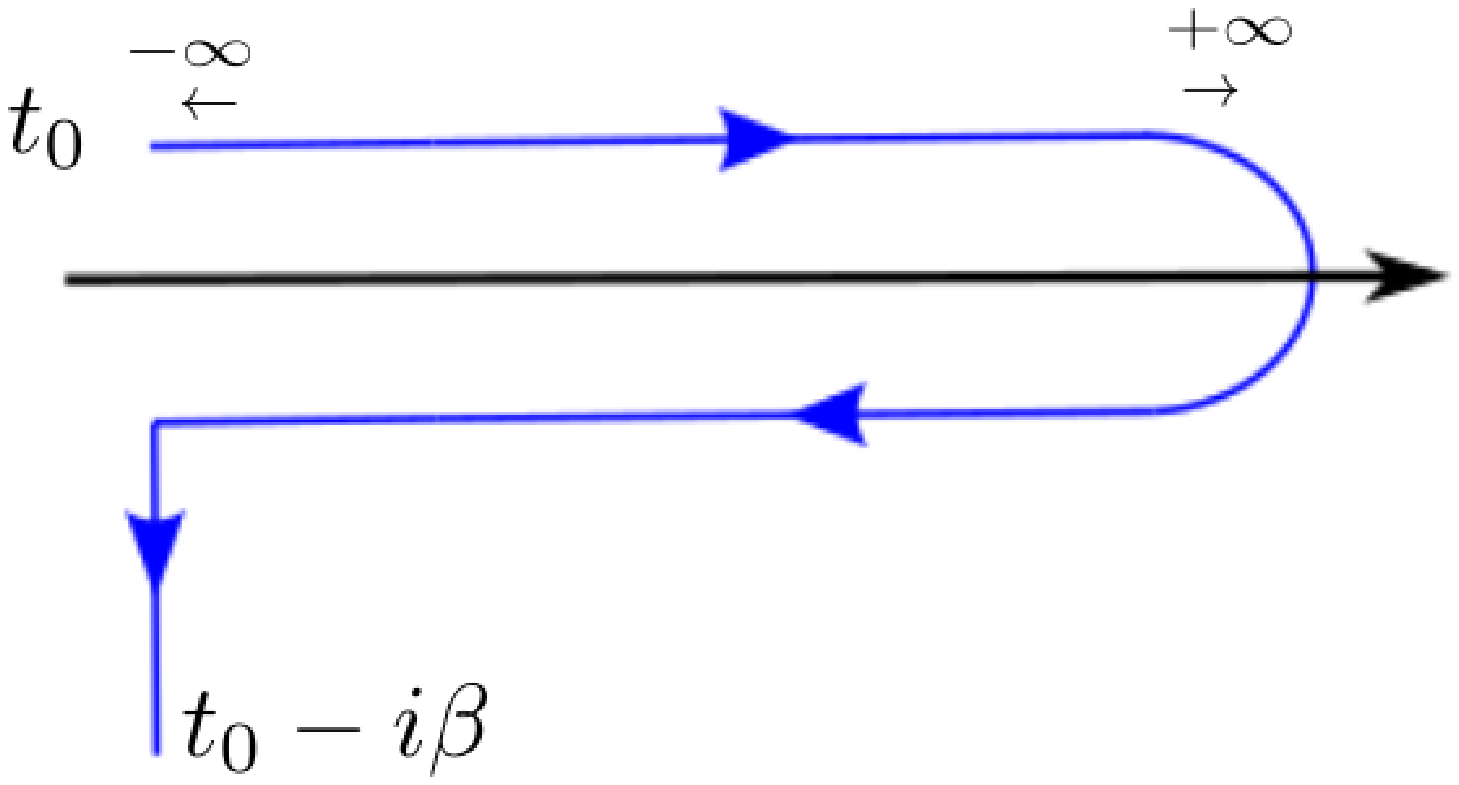}}\quad \quad
\subfigure[]{\includegraphics[scale=0.4]{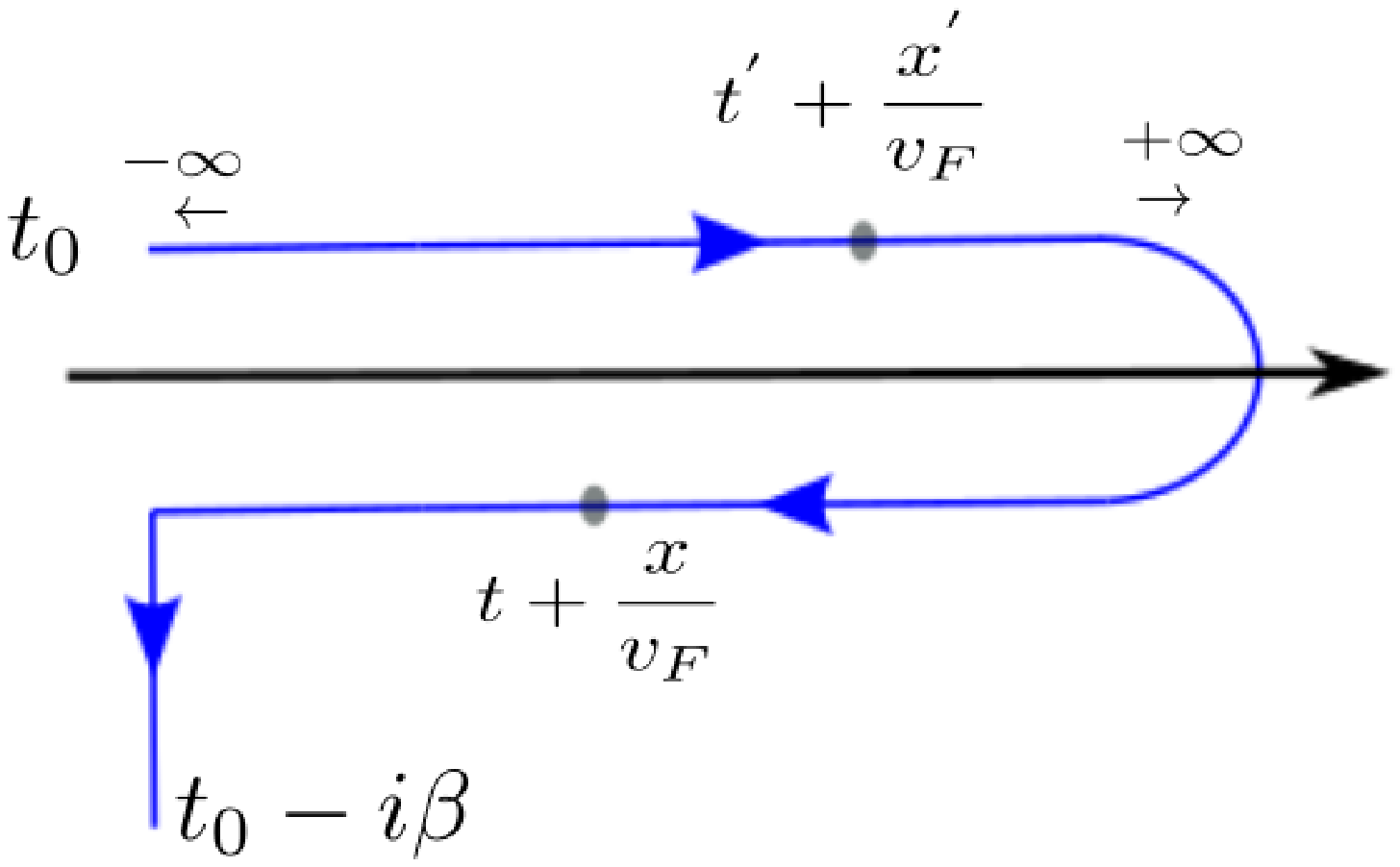}}\quad \quad
\subfigure[]{\includegraphics[scale=0.4]{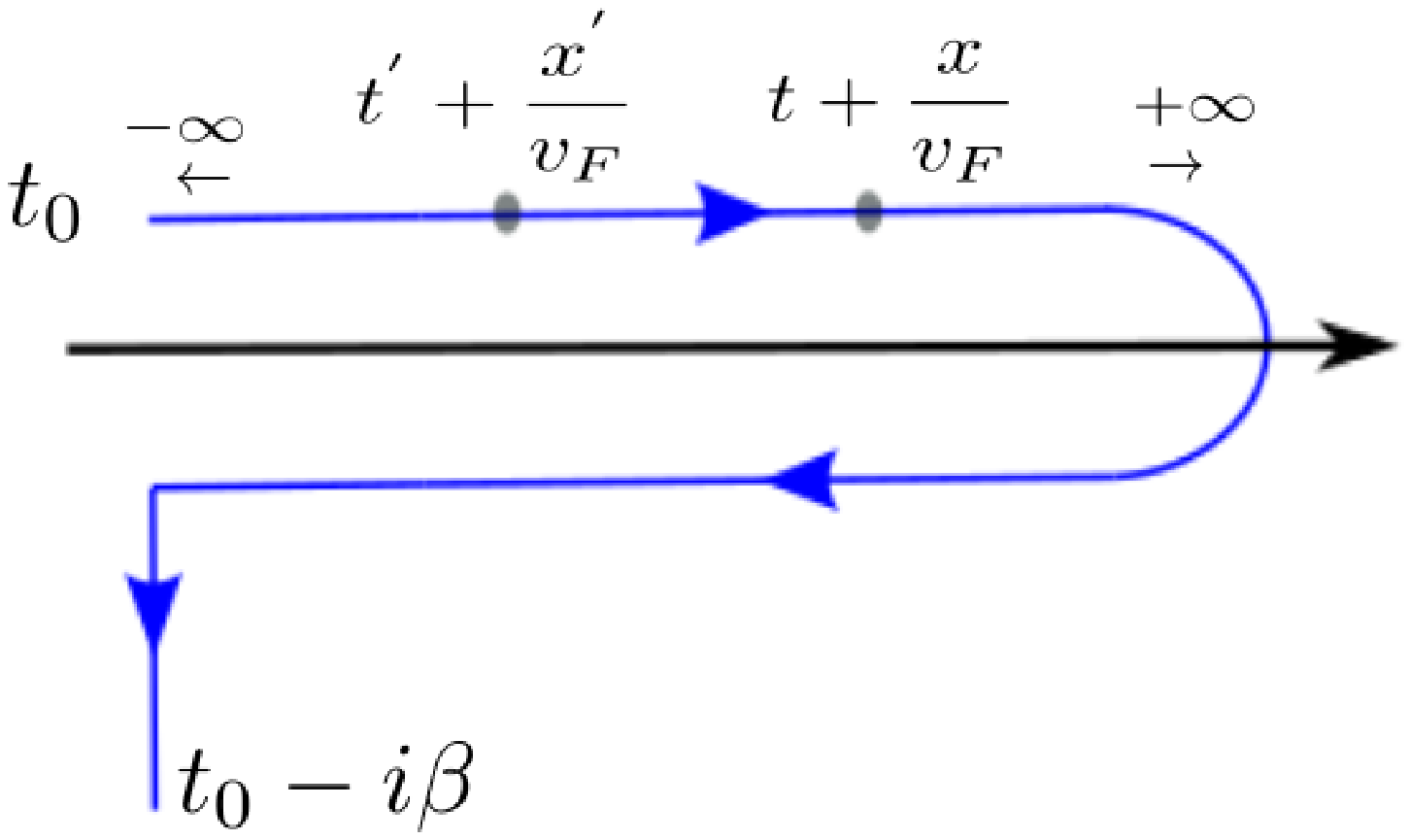}}
\caption{\small (\textbf{a}) The extended complex-time Keldysh contour on which Keldysh Green function theory is constructed. Times on the lower branch are greater than the upper branch. The contour is extended along the imaginary axis in a third branch to include the possibility of finite temperature Green functions. (\textbf{b}) $t$ is on the lower branch and $t^{'}$ is on the upper branch, so $t>t^{'}$ although on the real time axis it appears that $t^{'}>t$. (\textbf{c}) Both $t$ and $t^{'}$ are on the same branch of the contour.}
\label{figkeldysh}
\end{figure}

\setcounter{equation}{0}
\renewcommand{\theequation}{C.\arabic{equation}}
It is easy to reinterpret the space-time dependent Green functions of the main text as Keldysh Green functions where the times are on the Keldysh contour as shown in Fig.\ref{figkeldysh}.
For this we reinterpret $ U $ as follows,
\begin{align}
U(t,t^{'}) \mbox{          } = \mbox{            }
e^{ -i  \int_Cd\tau \mbox{      } \theta_C(t-\tau)\theta_C(\tau-t^{'})\mbox{            }e V_b(\tau) }
\end{align}
The standard meaning of the contour ordering $ \theta_C(t-t^{'}) $ is that $ \theta_C(t-t^{'}) = 1 $ if $ t $ is to the right of $ t^{'} $ on the contour $ C $ and $ \theta_C(t-t^{'}) = 0 $ if $ t $ is to the left of $ t^{'} $ on this contour and $ \theta_C(0) = \frac{1}{2} $ consistent with Dirichlet regularisation. The claim is that the Keldysh contour ordered Green function of the system may simply be written down as,
 \begin{align}
 <T_C\mbox{        } \psi_{\nu}(x,t)\psi^{\dagger}_{\nu^{'}}(x^{'},t^{'})> \mbox{ }=\mbox{ }\frac{i}{2\pi} \frac{\frac{\pi}{\beta v_{F}}}{\sinh( \frac{ \pi }{\beta v_F } (\nu x-\nu^{'}x^{'}-v_F(t-t^{'}) ) )} \kappa_{\nu,\nu^{'}}
 \end{align}
 where $\nu$,$\nu^{'} = \pm 1$.
The reader may recall from the main text that the quantities $ \kappa $ contains evolution functions such as $  U(t+ \frac{x}{v_F},t^{'}+\frac{x^{'}}{v_F}) $.
 For example if $ t $ is on the lower branch and $ t^{'} $ is on the upper branch of the contour and $ V_b(\tau) = 0 $ when $ \tau $ is purely imaginary then,
 \begin{align}
U(t+ \frac{x}{v_F}&,t^{'}+\frac{x^{'}}{v_F}) \mbox{          } = \mbox{            }
 e^{ -i  \int_Cd\tau \mbox{      } \theta_C(t + \frac{x}{v_F}
-\tau)\theta_C(\tau-t^{'}-\frac{x^{'}}{v_F})\mbox{            }e V_b(\tau) }
\end{align}
But since,
 \begin{align}
  \int_Cd\tau \mbox{      } \theta_C(t + \frac{x}{v_F}
-\tau)\theta_C(\tau-t^{'}-\frac{x^{'}}{v_F})\mbox{            }e V_b(\tau)
\mbox{          } = \mbox{            }
\int_{t^{'}+\frac{x^{'}}{v_F}}^{t + \frac{x}{v_F}}d\tau \mbox{      }  e V_b(\tau)
\end{align}
as before, we see that the reinterpretation has no effect on the results.
Similarly, if $ t,t^{'} $ are on the upper branch,
 \begin{align}
  \int_Cd\tau \mbox{      } \theta_C(t + \frac{x}{v_F}
-\tau)\theta_C(\tau-t^{'}-\frac{x^{'}}{v_F})\mbox{            }e V_b(\tau)
\mbox{          } = \mbox{            }   \theta(t + \frac{x}{v_F}-t^{'}-\frac{x^{'}}{v_F})
\int_{t^{'}+\frac{x^{'}}{v_F}}^{t + \frac{x}{v_F}}d\tau \mbox{      }  e V_b(\tau)
\end{align}
where $ \theta $ is now the ordinary Heaviside step function with a real argument.
Therefore reinterpreting the evolution function in the manner shown is sufficient to allow a recasting of the Green function in the main text as a Keldysh Green function where the time ordering is on the Keldysh contour.

\section*{APPENDIX D: Calculation of finite bandwidth tunneling current}
\label{AppendixD}
\setcounter{equation}{0}
\renewcommand{\theequation}{D.\arabic{equation}}
We already know from the derivation for the infinite bandwidth case the following expressions,
\begin{align}
c^{\infty}_{.,R}(t) = \left(  \sum_{p} c^{\infty}_{p,R}(t_0) e^{ -i (t-t_0)  p v_F  }\right)\mbox{      }  U(t_0,t)
  - i \Gamma  \int_{t_0}^t c^{\infty}_{.,L}(t_2)
  \delta(v_F(t-t_2))\mbox{ } U(t_2,t) \, dt_2
  \label{cinfr}
\end{align}
and
\begin{align}
c^{\infty}_{.,L}(t) = \left(  \sum_{p} c^{\infty}_{p,L}(t_0) e^{ i (t-t_0)  p v_F  }\right)
  - i \Gamma   \int_{t_0}^t c^{\infty}_{.,R}(t_2)
 \delta(v_F(t-t_2))  \, dt_2
 \label{cinfl}
\end{align}
Making use of the relation $c_{.,\nu}(t) =  c^{\infty}_{.,\nu}(t) + \delta c_{.,\nu}(t)$ and Eqs.\ref{cinfr}, \ref{cinfl}, \ref{cdotR} and \ref{cdotL} we obtain the following coupled equations
\begin{widetext}
\begin{align}
\delta c_{.,R}(t)\mbox{ }
 = \mbox{ } (&- \sum_{|p|>\Lambda}c^{\infty}_{p,R}(t_0) + \sum_p \delta c_{p,R}(t_0)) e^{ -i (t-t_0)  p v_F  }\mbox{      }  U(t_0,t)
  \nonumber \\&- i \Gamma  \int_{t_0}^t
 (\delta(v_F(t-t_2))\delta c_{.,L}(t) - c^{\infty}_{.,L}(t_2)  \Delta(v_F(t-t_2)))\mbox{ } U(t_2,t) \, dt_2
\end{align}
and
\begin{align}
\delta c_{.,L}(t) \mbox{ }=\mbox{ }   (&- \sum_{|p| > \Lambda}c^{\infty}_{p,L}(t_0) + \sum_p\delta c_{p,L}(t_0) )  e^{ i (t-t_0)  p v_F  }
  \nonumber \\ &- i \Gamma   \int_{t_0}^t
(\delta(v_F(t-t_2))\delta c_{.,R}(t_2) - \Delta(v_F(t-t_2))c^{\infty}_{.,R}(t_2)) \, dt_2
\end{align}
We solve the above two equations and write separate expressions for $\delta c_{.,R}(t)$ and $\delta c_{.,L}(t)$.
\begin{align}
\delta c_{.,R}(t)\mbox{ } = \mbox{     }  &\frac{2 v_F }{\Gamma ^2+4v_F^2}(2v_F ((- \sum_{|p|>\Lambda}c^{\infty}_{p,R}(t_0) + \sum_p \delta c_{p,R}(t_0)) e^{ -i (t-t_0)  p v_F  }\mbox{      }  U(t_0,t) \nonumber \\ &+ i \Gamma
  \int_{t_0}^t c^{\infty}_{.,L}(t_2)  \Delta(v_F(t-t_2))\mbox{ } U(t_2,t) \, dt_2)-i \Gamma  (    (- \sum_{|p| > \Lambda}c^{\infty}_{p,L}(t_0)
  + \sum_p\delta c_{p,L}(t_0) )  e^{ i (t-t_0)  p v_F  } \nonumber \\
  &+ i \Gamma   \int_{t_0}^t
 \Delta(v_F(t-t_2))c^{\infty}_{.,R}(t_2) \, dt_2))
 \label{deltacr}
\end{align}
and
\begin{align}
\delta c_{.,L}(t)\mbox{ } = \mbox{     }&\frac{2 v_F }{\Gamma ^2+4v_F^2}(2v_F (   (- \sum_{|p| > \Lambda}c^{\infty}_{p,L}(t_0) + \sum_p\delta c_{p,L}(t_0) )  e^{ i (t-t_0)  p v_F  } \nonumber \\
&+ i \Gamma   \int_{t_0}^t
 \Delta(v_F(t-t_2))c^{\infty}_{.,R}(t_2) \, dt_2) -i \Gamma  ((- \sum_{|p|>\Lambda}c^{\infty}_{p,R}(t_0) + \sum_p \delta c_{p,R}(t_0)) e^{ -i (t-t_0)  p v_F  }\mbox{      }  U(t_0,t) \nonumber \\
 &+ i \Gamma
  \int_{t_0}^t c^{\infty}_{.,L}(t_2)  \Delta(v_F(t-t_2))\mbox{ } U(t_2,t) \, dt_2) )
  \label{deltacl}
\end{align}
 Also Eqs.\ref{cinfr} and \ref{cinfl} reduce to
\begin{align}
c^{\infty}_{.,R}(t)\mbox{   } = \mbox{    }  \frac{2 v_F (2 v_F  ( \sum_p c^{\infty}_{p,R}(t_0) e^{ -i (t-t_0)  p v_F  } U(t_0,t))
 -i \Gamma  (\sum_p c^{\infty}_{p,L}(t_0) e^{ i (t-t_0)  p v_F  }))}{\Gamma ^2+4 v_F^2}
 \label{cinfrfinal}
\end{align}
and
\begin{align}
c^{\infty}_{.,L}(t)\mbox{   } = \mbox{    } \frac{2 \left(2  v_F^2(\sum_p c^{\infty}_{p,L}(t_0) e^{ i (t-t_0)  p v_F  })-i \Gamma v_F  ( \sum_p c^{\infty}_{p,R}(t_0) e^{ -i (t-t_0)  p v_F  } U(t_0,t)) \right)}{\Gamma ^2+4 v_F^2}
\label{cinflfinal}
\end{align}
\end{widetext}
Using Eqs. \ref{deltacr}, \ref{deltacl}, \ref{cinfrfinal} and \ref{cinflfinal} and the corresponding complex conjugates we can write down expressions for the correlations of the type $<\delta c^{\dagger}_{.,\nu^{'}}(t^{'})c^{\infty}_{.,\nu}(t)>$ and $<c^{\dagger \infty}_{.,\nu^{'}}(t^{'})\delta c_{.,\nu}(t)>$. After some simplification these correlation functions are obtained as some non-trivial combinations of the bias $V_{b}(t)$ and the equal-time equilibrium infinite bandwidth Green functions which we already know but for $<T\mbox{       } \delta \psi_{R}(x,t_{0})\psi^{\dagger,\infty}_{\nu^{'}}(x^{'},t_{0}) >_{eq} $ and $<T\mbox{       } \delta \psi_{L}(x,t_{0})\psi^{\dagger,\infty}_{\nu^{'}}(x^{'},t_{0}) >_{eq}$ (where $\nu^{'} = R,L$) which we are required to explicitly calculate. Note that the Green functions at equal-time $t_{0}$ implies equilibrium Green functions as we take $t_{0} \rightarrow -\infty$ i.e. long before the bias is switched on.
\begin{widetext}
In equilibrium we have
\begin{align}
&i \partial_t \delta c_{p,R}(t) = v_Fp \mbox{    }  \delta c_{p,R}(t) + \frac{\Gamma}{L}\delta c_{.,L}(t)- \theta(|p|-\Lambda)\frac{\Gamma}{L}c^{\infty}_{.,L}(t) \nonumber \\&
i \partial_t \delta c_{p,L}(t) = -v_Fp \mbox{     } \delta c_{p,L}(t) + \frac{\Gamma}{L} \delta c_{.,R}(t)- \theta(|p|-\Lambda)\frac{\Gamma}{L}c^{\infty}_{.,R}(t)
\end{align}
We transform from time to discrete Matsubara frequency ($z_{n} = \frac{(2n+1) \pi}{\beta}$) and write down the correlations,
\begin{align}
<T\mbox{       } \delta c_{p,R}(n)c^{\dagger,\infty}_{p^{'},\nu^{'}}(n) >
 = - \theta(|p|-\Lambda)\frac{\Gamma}{L}\frac{1}{(i z_n-v_Fp) }<T\mbox{       } c^{\infty}_{.,L}(n)c^{\dagger,\infty}_{p^{'},\nu^{'}}(n) >
  + \frac{\Gamma}{L}\frac{1}{(i z_n-v_Fp) }<T\mbox{       } \delta c_{.,L}(n)c^{\dagger,\infty}_{p^{'},\nu^{'}}(n) >
  \end{align}
  and
  \begin{align}
  <T\mbox{       } \delta c_{p,L}(n)c^{\dagger,\infty}_{p^{'},\nu^{'}}(n) >
 = - \theta(|p|-\Lambda)\frac{\Gamma}{L}\frac{1}{(i z_n + v_Fp) }<T\mbox{       } c^{\infty}_{.,R}(n)c^{\dagger,\infty}_{p^{'},\nu^{'}}(n) >
  +  \frac{\Gamma}{L} \frac{1}{(i z_n + v_Fp) }<T\mbox{       } \delta c_{.,R}(n)c^{\dagger,\infty}_{p^{'},\nu^{'}}(n) >
  \end{align}
  After some algebra we Fourier transform to real space which allows for further simplification. We then transform the discrete frequencies back to time taking the time interval $t-t^{'}$ to be small. Finally we obtain the following correlations for small $t-t^{'}$ and large finite bandwidth
  \begin{align}
  <T\mbox{       } \delta \psi_{R}(x,t)\psi^{\dagger,\infty}_{\nu^{'}}(x^{'},t^{'}) >_{eq}
 \mbox{ }\approx\mbox{ } 4 i \Gamma  v_F
   \Gamma
 \frac{ (2 v_F)^2 }{ (\Gamma ^2+4 v_F^2)^2 } \frac{ i }{ -i\beta }
 \delta_{\nu^{'},-1} \mbox{ }
 \frac{i \Gamma   \theta (x x^{'}) \coth \left(\frac{\pi  (x+x^{'})}{\beta  v_F}\right) \text{csch}\left(\frac{\pi  (x+x^{'})}{\beta  v_F}\right)}{2 \beta  \Lambda  v_F^4} \nonumber \\ + ( \Gamma^2 - (2v_F)^2 ) \mbox{  }  \Gamma
 \frac{ (2 v_F)^2 }{ (\Gamma ^2+4 v_F^2)^2 }\frac{ i }{ -i\beta }
\delta_{\nu^{'},1}
\mbox{ }
\frac{1}{L^2}
  \mbox{     } \mbox{   }
\frac{i \Gamma  L^2 (\text{sgn}(x) \theta (-x x^{'})) \coth \left(\frac{\pi  (x^{'}-x)}{\beta v_F}\right)
\text{csch}\left(\frac{\pi  (x^{'}-x)}{\beta  v_F}\right)}{2 \beta  \Lambda  v_F^4}
\label{deltapsipsir}
 \end{align}
 and
 \begin{align*}
 <T\mbox{       } \delta \psi_{L}(x,t)\psi^{\dagger,\infty}_{\nu^{'}}(x^{'},t^{'}) >_{eq}
 \mbox{     }   \approx\mbox{     }
 \end{align*}
 \begin{align}
 -\frac{\Gamma}{L^2}  \mbox{     }
\mbox{     } \frac{  iL }{  v_F }
\mbox{       } \frac{ (2 v_F)^2 }{ (\Gamma ^2+4 v_F^2)^2 } \frac{i \Gamma     }{\pi  \Lambda  v_F^2} \frac{ i }{ -i\beta }
 4 i  v_F  \Gamma  \delta_{\nu^{'},1}
\mbox{     }
   \theta( x x^{'}) \mbox{     } \frac{  iL }{  v_F }
  \mbox{    }
 \left(  \frac{\pi  \coth \left(\frac{\pi  (x+ x^{'} )}{\beta  v_F}\right) \text{csch}\left(\frac{\pi  (x+ x^{'})}{\beta v_F}\right)}{2 \beta } \right) \nonumber \\
 -   \frac{\Gamma}{L^2}  \mbox{     }
\mbox{     } \frac{  iL }{  v_F }
\mbox{       } \frac{ (2 v_F)^2 }{ (\Gamma ^2+4 v_F^2)^2 } \frac{i \Gamma   }{\pi  \Lambda  v_F^2} \frac{ i }{ -i\beta }
 (\Gamma^2 -(2v_F)^2) \delta_{\nu^{'},-1}  \mbox{ }
  \mbox{     }
(  \text{sgn}(x^{'}) \theta (-x x^{'})) \mbox{     } \frac{  iL }{  v_F }   \left(  \frac{\pi  \coth \left(\frac{\pi  (x-x^{'})}{\beta  v_F}\right) \text{csch}\left(\frac{\pi  (x-x^{'})}{\beta  v_F}\right)}{2 \beta }\right)
\label{deltapsipsil}
 \end{align}
\end{widetext}
Now we have all the ingredients to calculate $<\delta c^{\dagger}_{.,\nu^{'}}(t^{'})c^{\infty}_{.,\nu}(t)>$ and $<c^{\dagger \infty}_{.,\nu^{'}}(t^{'})\delta c_{.,\nu}(t)>$. Substituting in Eq.\ref{ditun} and simplifying we obtain Eq.\ref{deltaifinal} in the main text.

\newpage

\section*{References}
\bibliographystyle{iopart-num}
\bibliography{ref}
\end{document}